\documentclass[12pt]{article} 

\usepackage{fullpage}
\usepackage{amssymb,amsmath}
\usepackage{xspace}
\usepackage{graphicx}
\usepackage{amssymb}
\usepackage{amsmath}
\usepackage{multicol}
\usepackage{ulem}
\usepackage{hyperref}
\usepackage{caption}
\usepackage{subcaption}
\usepackage{float}

\newcommand{\sk}{Skyrmion\xspace}
\newcommand{\sks}{Skyrmions\xspace}
\newcommand{\bx}{\mbox{\boldmath $x$}}
\newcommand{\bpi}{\mbox{\boldmath $\pi$}}

\newcommand{\btau}{\mbox{\boldmath $\tau$}}
\newcommand{\fg}{figure\xspace}
\newcommand{\fig}{figure\xspace}
\newcommand{\Fig}{Figure\xspace}
\newcommand{\Figs}{Figures\xspace}

\newcommand{\fgs}{figures\xspace}
\newcommand{\bd}{baryon-density\xspace}
\newcommand{\bdn}{baryon density\xspace}

\begin{document}
\title{
  \vskip 15pt
  {\bf \Large \bf Scattering of Skyrmions}
  \vskip 10pt}
\author{David Foster\thanks{dave.foster@bristol.ac.uk}~\\[5pt]
{\normalsize {\sl HH Wills Physics Laboratory,}}\\
{\normalsize {\sl Tyndall Avenue, Bristol BS8 1TL}}\\
{\normalsize {\sl United Kingdom}}\\[5pt]
and Steffen Krusch\thanks{S.Krusch@kent.ac.uk } \\[5pt]
{\normalsize {\sl School of Mathematics, Statistics and Actuarial 
Science}}\\
{\normalsize {\sl University of Kent,
Canterbury CT2 7NF}}\\
{\normalsize {\sl United Kingdom}}
}

\date{\today}
\maketitle

\begin{abstract}
In this paper, we present a detailed study of Skyrmion-Skyrmion scattering for two $B=1$ Skyrmions in the attractive channel where we observe two different scattering regimes. For large separation, the scattering can be approximated as interacting dipoles. We give a qualitative estimate when this approximation breaks down. For small separations we observe an additional short-range repulsion which is qualitatively similar to monopole scattering. We also observe the interesting effect of ``rotation without rotating'' whereby two Skyrmions, whose orientations remain constant while well-separated, change their orientation after scattering. We can explain this effect by following preimages through the scattering process, thereby measuring which part of an in-coming Skyrmion forms part of an out-going Skyrmion. This leads to a new way of visualising Skyrmions. Furthermore, we consider spinning Skyrmions and find interesting trajectories. 
\end{abstract}

\vspace{0.5cm} 
\centerline{PACS-number: 12.39.Dc}
\vspace{0.5cm}
\begin{tt}
\pageref{lastref} pages, 16 figures
\end{tt}

\newpage

\section{Introduction}

The Skyrme model is a nonlinear field theory model of atomic nuclei \cite{Skyrme}. As a classical field theory, this model has soliton solutions, known as \sks, which are stabilised by a conserved topological charge. Skyrmions have been calculated for various charges, see e.g. \cite{Battye:2001qn} for a comprehensive summary, and \cite{Battye:2004rw, Battye:2006tb,Feist:2012ps} for more recent results when it became apparent that massive pions play an important role. When these Skyrmions are quantised, as fermions, they model protons and neutrons \cite{Adkins:1983ya, Adkins:1983hy}. An important ingredient in the quantisation are the so-called Finkelstein-Rubinstein constraints \cite{Finkelstein:1968hy}, which guarantee that Skyrmions can be consistently quantised as fermions. Using the symmetries of classical Skyrmions, these constraints also allow the quantum numbers of the ground and excited states to be calculated \cite{Irwin:1998bs, Krusch:2002by, Krusch:2005iq}.
Reference \cite{Battye:2009ad} included massive pions and found that the energies of quantum ground and excited states of Skyrmions had good qualitative and reasonable quantitive agreement with experimental results, for even topological charges.
However, the approach does not produce good results for odd values of the  topological charge greater than three. This may be related to the fact that Skyrmions deform when they are spinning \cite{Battye:2005nx} or isospinning \cite{Battye:2014qva}.  More recently, properties of Carbon-12 have been successfully modelled using the Skyrme model \cite{Lau:2014baa}. These calculations helped to understand the structure of the ground state of Carbon-12 and the so-called Hoyle state.

In nuclear physics, scattering experiments are very important. However, relatively little progress has been made with Skyrmion-Skyrmion scattering, and its applications to nuclear physics. Classical Skyrmion scattering was first discussed using an axially-symmetric approximation in \cite{Verbaarschot:1986rj}. The first numerical full field  simulation of Skyrmion scattering for two $B=1$ Skyrmions was performed in \cite{Allder:1987kq}. Skrymion scattering for different charges with symmetric initial conditions was discussed in \cite{Battye:1996nt}. The similarity with monopole scattering led to various important developments \cite{Manton} including the rational map ansatz \cite{Houghton:1997kg}.  
From a more analytical point of view, Manton discussed low energy Skyrmion scattering using the idea of an unstable manifold \cite{Manton:1988ba, Gisiger:1994gj} and the geodesic approximation \cite{Manton:1981mp}. This unstable manifold can be mapped out exactly for well-separated Skyrmions \cite{Irwin:1996nj} and has been calculated numerically in \cite{Waindzoch:1997rq}. Schroers discussed the interaction of well-separated moving and spinning Skyrmions \cite{Schroers:1993yk}, see also \cite{Gisiger:1994gj, Gisiger:1994pr} for related results. 
Braaten discussed how to calculate scattering cross sections from the Skyrme model \cite{Braaten:1987hk}. Gisiger and Paranjape presented a comprehensive, pedagogical introduction to these ideas and calculated an analytic approximation to low energy nucleon-nucleon scattering \cite{Gisiger:1998tv}.

In this paper, we focus on classical scattering of two charge one Skyrmions with variable impact parameter. The paper is organised as follows. In section \ref{SkyrmeModel} we review the Skyrme model with a particular emphasis on the dipole interaction.  In section \ref{SkyrmionScattering} we present a numerical study of Skyrmion scattering.
We then describe Skyrmion-Skyrmion scattering in the attractive channel using the classical dipole approximation. We also derive the dynamics in the relativistic case and discuss the modifications for nonzero pion mass. We observe the interesting effect of ``rotation without rotating''. In section 
\ref{Visualisation} we introduce a new way of visualising Skyrmions which  explains this effect. We then discuss scattering of two spinning Skyrmions. In section \ref{Monopoles} we give a brief comparison of monopole and Skyrmion scattering. We end with a conclusion and discuss open problems.

\section{The Skyrme model}
\label{SkyrmeModel}

The Skyrme model is a three dimensional non-linear theory of pions where the field $U(t,\bx)$ is an $\mbox{SU}(2)$-valued scalar. It is a low energy effective theory of QCD and is defined by the Lagrangian \cite{Manton},
\begin{equation}
\label{Lag}
L=\int\left\{-\frac{1}{2}\mbox{Tr}\,(R_\mu R^\mu)+
\frac{1}{16}\mbox{Tr}([R_\mu,R_\nu][R^\mu,R^\nu]) - {m_\pi}^2 \mbox{Tr} (1_2-U)
\right\}\mbox{d}^3x,
\end{equation}
where $R_\mu = \partial_\mu U U^\dagger,$ $1_2$ is the unit matrix in two dimensions and $m_\pi$ parametrises the pion mass. Here we have expressed the model in so-called Skyrme units, where we have chosen an energy unit $\frac{F_\pi}{4e}$ and a length unit $\frac{2}{eF_\pi}$. $F_\pi$ is the pion decay constant and $e$ is a dimensionless parameter. Field configurations can only have finite energy provided that the field $U({\bf x},t) \to 1_2$ as $|{\bf x}| \to \infty.$ Hence, finite-energy fields are defined on the one-point compactification of ${\mathbb R}^3$, namely ${\mathbb R}^3 \cup \{\infty\} \cong S^3.$ Furthermore, the target space $SU(2)$ is homeomorphic to $S^3.$ Therefore, finite-energy configurations belong to an element of the third homotopy group $\pi_3(S^3)\cong \mathbb{Z}$ and are indexed by an integer. This integer is the topological charge, $B$, and is interpreted as the baryon number. In atomic nuclei, $B$ corresponds to the sum of the number protons and neutrons. The topological charge can be calculated as an integral over the baryon density ${\cal B}(\bx)$ namely,
\begin{equation}
\label{Bd}
B = \int_{{\mathbb R}^3} {\cal B}(\bx) \mbox{d}^3x, \quad {\rm where} \quad
{\cal B}(\bx) = -\frac{\epsilon_{ijk}}{24 \pi^2} \mbox{Tr} \left(R_i R_j R_k\right).
\end{equation}
It is often more convenient to reparameterise the Skyrme field with three pion fields $\bpi=(\pi_1,\pi_2,\pi_3)^{T}$ and a constrained field $\sigma$ as $U=\sigma 1_2+i\bpi\cdot\btau$, where $\sigma^2+\bpi\cdot\bpi=1$ and $\btau$ is the triplet of Pauli matrices. We shall be making use of this later. Numerical evidence suggests that the $B=1$ Skyrmion is spherically symmetric. This is best described with the so-called hedgehog ansatz,
\begin{equation}
U_{\rm H} = \cos f(r)\, 1_2 +i \sin f(r) \hat{\bx} \cdot \btau, \label{hog}
\end{equation}
where $r=|{\bf x}|$ and ${\hat{\bf x}} = {\bf x}/r.$ For minimum-energy solutions the shape function $f(r)$ has to be calculated numerically subject to the boundary conditions $f(0) = \pi$ and $f(\infty)=0.$

For massless pions, $m_\pi=0,$ the interaction of two well-separated $B=1$ Skyrmions can be approximated by the dipole-dipole interaction \cite{Manton}
\begin{equation}
\label{Eint0}
E_{{\rm int}} = -\frac{2C^2}{3 \pi} \left(1-\cos \psi\right)
\frac{1-3\left({\bf {\hat X}} \cdot {\bf {\hat n}}\right)^2}{X^3},
\end{equation}
where $C$ is the dipole strength, and $\psi$ is the angle one of the Skyrmions is rotated through about the axis given by the unit vector ${\bf {\hat n}}$. The vector ${\bf X}$ is the difference between the position vectors of the two Skyrmions,  $X = |{\bf X}|$ is their separation and ${\bf {\hat X}} = {\bf X}/X.$ For a $B=1$ Skyrmion the constant $C$ is given by $C=2.16$ ($m_\pi=0$) \cite{Manton}. The value of $C$ corresponds to the leading order term in the large $r$ expansion of the shape function $f(r) \sim \frac{C}{r^2}$. This can be shown by linearising the equations of motion for $f(r)$. In this paper we are only interested in when the interaction energy \eqref{Eint0} is minimal, namely when $\psi = \pi$ and ${\bf {\hat X}} \cdot {\bf {\hat n}} = 0.$ We define this as the attractive channel, and the interaction energy simplifies to
\begin{equation}
\label{Eint0a}
E_{{\rm int}}^{{\rm att}} = -\frac{4C^2}{3 \pi} 
\frac{1}{X^3}.
\end{equation}

As a point of notation we define the \sk locations as the points in $\mathbb{R}^3$ where $U(\bx)=-1_2, (\sigma=-1, \pi_a=0)$. This is the antipodal point of the vacuum and is hence a region of large energy density for the $B=1$ hedgehog configuration \eqref{hog}.

\section{Skyrmion Scattering} \label{SkyrmionScattering}

In this article we are investigating \sk scattering. There has been some analytical progress using the instanton ansatz \cite{Atiyah:1989dq, Atiyah:1992if}, but so far the most productive method is to use numerical simulations. 

To achieve this we first need an initial configuration to evolve. We create a suitable configuration by numerically solving the equations of motion for the hedgehog ansatz, for the value of $m_\pi$ which we are interested in. This gives us a shapefunction $f(r)$ for the single \sk. We use this with the hedgehog ansatz and the product ansatz,
 \begin{equation}
 U(t,x,y,z)=U_1(\gamma(x-vt), y,z) U_2(\gamma(x+vt), y,z),
 \end{equation}
to give a two-\sk initial configuration. Here, $U_1$ is the hedgehog solution $U_1=U_{\rm H}(x+\frac{D}{2},y+\frac{b}{2},z)$ and $U_2$ is the hedgehog solution $U_2=\tau_3 U_{\rm H}(x-\frac{D}{2},y-\frac{b}{2},z) \tau_3$, which has been rotated by $\pi$ about the $z$-axis in target space by the $SU(2)$ matrix $\tau_3.$ This isorotation ensures that the \sks are in the attractive channel. Here, $\gamma =1/\sqrt{1-v^2}$ is the usual Lorentz factor. 

Throughout this paper we consistently chose the hedgehog ansatz \eqref{hog} to be orientated such that under $z \mapsto -z$, $\pi_3 \mapsto -\pi_3$.
We then evolved this initial configuration using a finite difference
leap-frog method on a discretised regular lattice. We chose a lattice
spacing of $\delta_x =0.1$ with either $100$ lattice points or $120$ lattice
points for large $b$. Therefore, $x,y$ and $z$ had the ranges $(-5,5)$ or
$(-6,6)$, depending on the number of lattice points. To minimise the effects
of radiation, and to replicate the infinite plane, we damped the boundary of
the box by smoothly introducing an extra ${\dot{U}}$ term in the equations
of motion at the boundary. This term damped the radiation and reduced the
reflection off the boundary. We chose to use leap-frog as it is a symplectic integrator, and we argue that preserving momentum is very important during a scattering process. 

\subsection{Numerical Results}

\begin{figure}[!ht]
\begin{center}
\includegraphics[width=4.5cm]{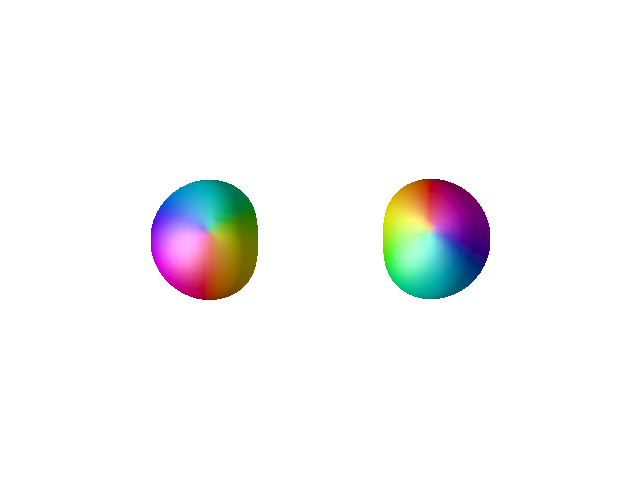}
\includegraphics[width=4.5cm]{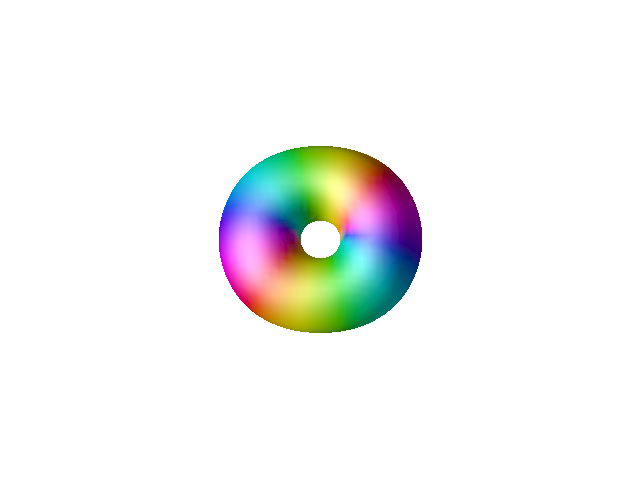}
\includegraphics[width=4.5cm]{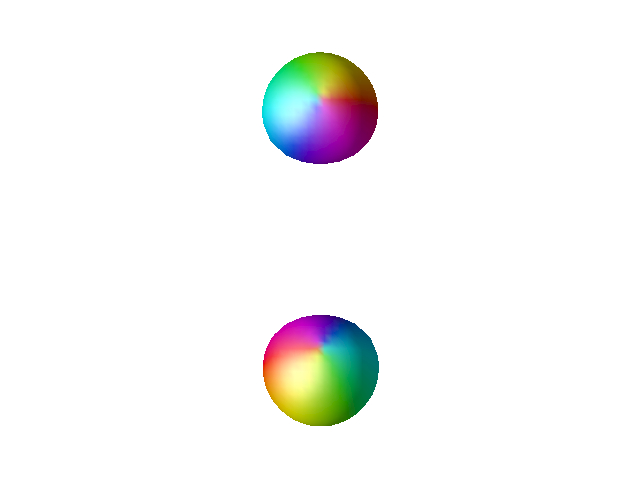}
\includegraphics[width=4.5cm]{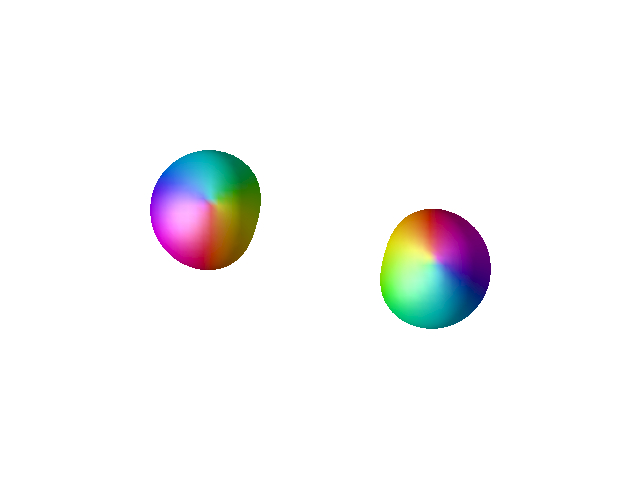}
\includegraphics[width=4.5cm]{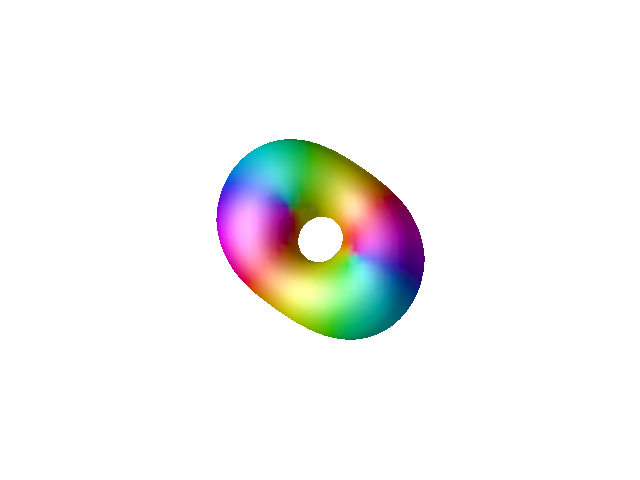}
\includegraphics[width=4.5cm]{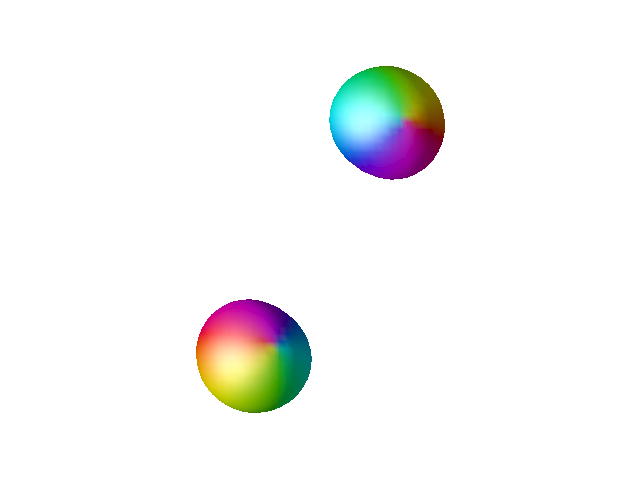}
\caption{Skyrmion scattering plots for $m_\pi=0$ and $v=0.2.$ Each row displays the initial, intermediate and final configuration. In the first row the impact parameter is $b=0,$ in the second row $b=0.4.$
\label{Fig5}}
\end{center}
\end{figure}

\begin{figure}[!ht]
\begin{center}
\includegraphics[width=4.5cm]{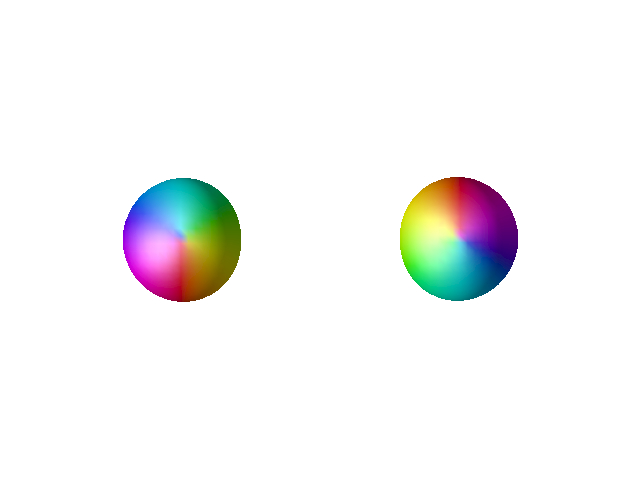}
\includegraphics[width=4.5cm]{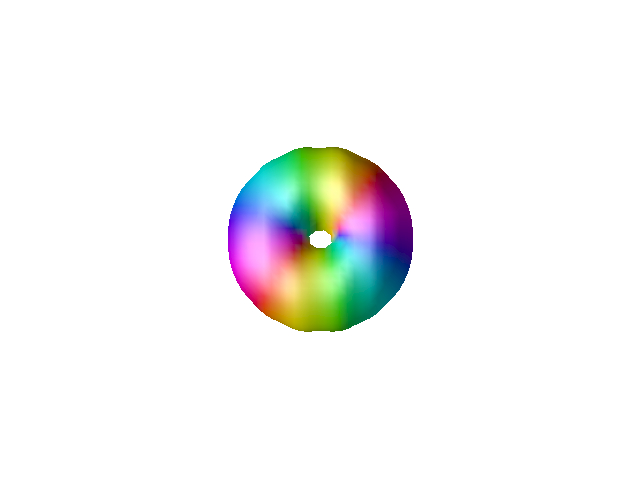}
\includegraphics[width=4.5cm]{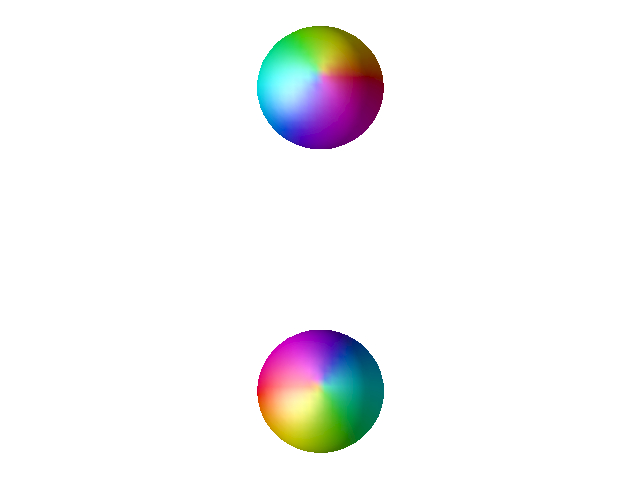}
\includegraphics[width=4.5cm]{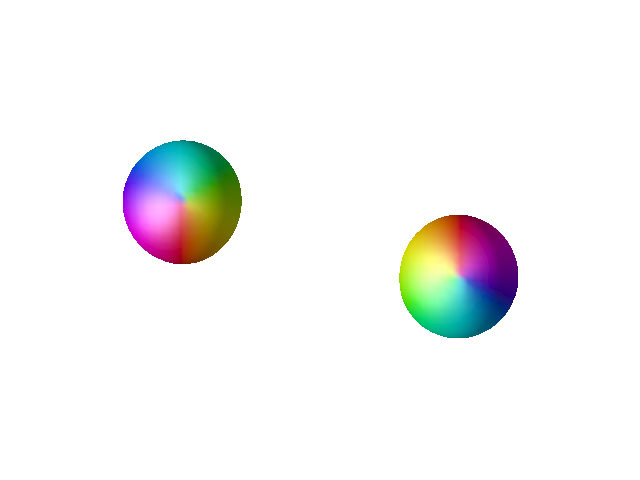}
\includegraphics[width=4.5cm]{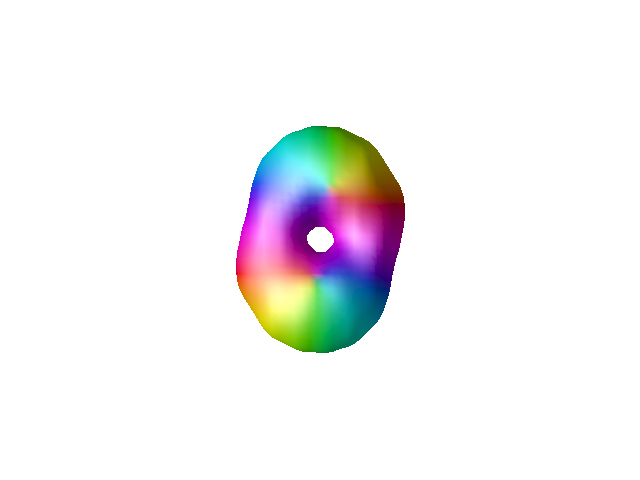}
\includegraphics[width=4.5cm]{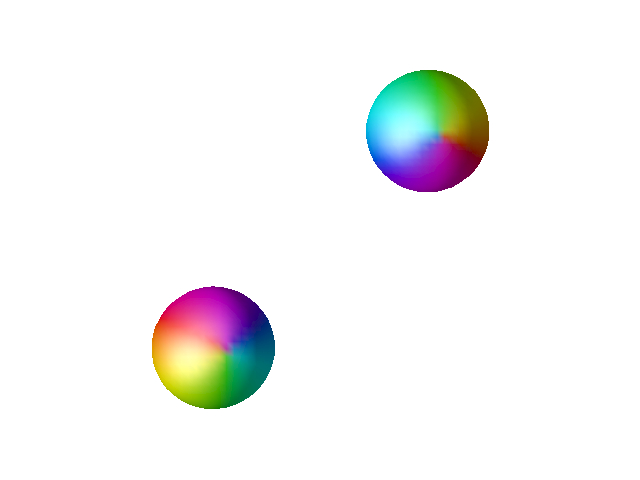}
\caption{Skyrmion scattering plots for $m_\pi=1$ and $v=0.2.$ Each row displays the initial, intermediate and final configuration. In the first row the impact parameter is $b=0,$ in the second row $b=0.4.$
\label{Fig6}}
\end{center}
\end{figure}

In figures \ref{Fig5} and \ref{Fig6} we display snapshots of the scattering of two $B=1$ Skyrmions. Throughout the text we colour the \sk \bd level-set plots to show the angle the pion fields have from  the $\hat{\pi}_2$-axis on the $\hat{\pi}_1,\hat{\pi}_2$ plane. It is coloured such that when the field lies  slightly above the $\hat{\pi}_2$-axis the colour is orange and when it is slightly below the colour is red. There is a detailed discussion of this colouring scheme and its physical interpretation in \cite{Manton:2011mi}.

In figure \ref{Fig5} the top row shows Skyrmion scattering for $m_\pi=0$ and zero impact parameter, $b=0$, with initial speed $v=0.2.$ The initial configuration is on the left. With the colouring scheme it is easy to see that the second Skyrmion is rotated by $\pi$ around the $z$-axis. The Skyrmions keep their orientation even as they merge and form the torus. However, when they reemerge as individual Skyrmions after passing through the torus configuration their orientation has changed. This is a rather intriguing effect of changing orientation without actually rotating. We discuss this phenomenon further in section \ref{preimages}.

Figure \ref{Fig6} shows the same set of snapshots but for $m_\pi=1.$ In the initial configuration the Skyrmions are more spherical, since the interaction force is weaker, leading to less deformation. The torus in the intermediate configuration is more compact with a smaller hole as expected for massive Skyrmions, see \cite{Foster:2013bw} for a detailed discussion. 

\begin{figure}[!htb]
       \centering
      \begin{subfigure}[b]{0.5\textwidth}
               \centering
    \includegraphics[width=\textwidth]{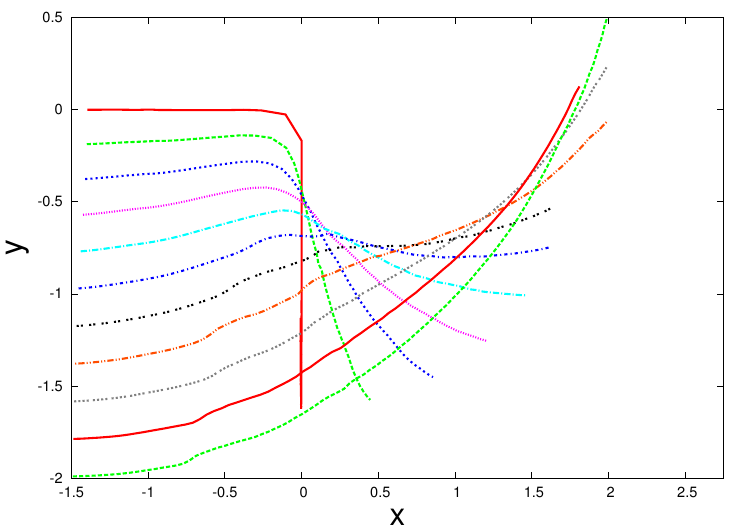}
               \caption{$m_\pi=0$}
               \label{m0-v0pt2-trajectories-one-side} 
       \end{subfigure}%
        ~ 
        \begin{subfigure}[b]{0.5\textwidth}
                \centering
  \includegraphics[width=\textwidth]{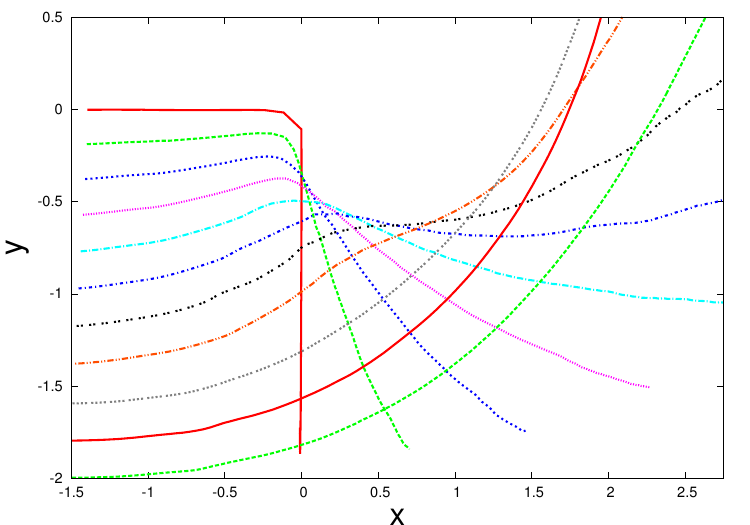}
                \caption{$m_\pi=0.5$}
                \label{m0pt5-v0pt2-trajectories-one-side}
        \end{subfigure}
        ~ 
        \caption{Trajectories of the location of a single \sk throughout a
scattering process.} \label{Trajs}
\end{figure}

Throughout the numerical simulation we tracked the \sk locations and to increase accuracy we
interpolated field values in-between lattice points. This gives the curves
in \fgs \ref{m0-v0pt2-trajectories-one-side} and
\ref{m0pt5-v0pt2-trajectories-one-side} which show the trajectories of the
location in the scattering plane.
These images show how the pion mass, $m_\pi$, affects the scattering process. For example it can be seen that for large separation the \sk with $m_\pi=0.5$ is deflected less.

\subsection{Dipole approximation}
As discussed earlier, for $m_\pi=0$, the attractive channel has the interaction energy \eqref{Eint0a}
$$
E_{{\rm int}}^{{\rm att}} = -\frac{4C^2}{3 \pi} 
\frac{1}{X^3}.
$$
%
For simplicity, we firstly describe the non-relativistic dynamics. Two $B=1$ Skyrmions can be approximated as point particles of mass $M \approx 1.232$, which is the rest mass of a single $B=1$ Skyrmion. We can then separate off the centre of mass motion, and the equations of motion can be written in terms of the relative coordinate ${\bf X}$ as  
\begin{equation}
\label{eom}
\mu {\ddot{\bf X}} = -\nabla E_{{\rm int}}^{{\rm att}},
\end{equation}
where $\mu = M/2$ is the reduced mass. Note that in the attractive channel the force between the Skyrmions is a central force, hence the relative angular momentum 
\begin{equation}
\label{lrel}
{\bf l}_{{\rm rel}} = \mu {\bf X} \times {\dot {\bf X}} 
\end{equation}
is conserved, and the dynamics takes place in a plane orthogonal to ${\bf n}.$
This two dimensional plane contains the non-trivial dynamics in the attractive channel and is known as the scattering plane. In the following we choose coordinates such that the scattering plane is given by $z=0.$

\begin{figure}[!ht]
\begin{center}
\includegraphics[width=10cm]{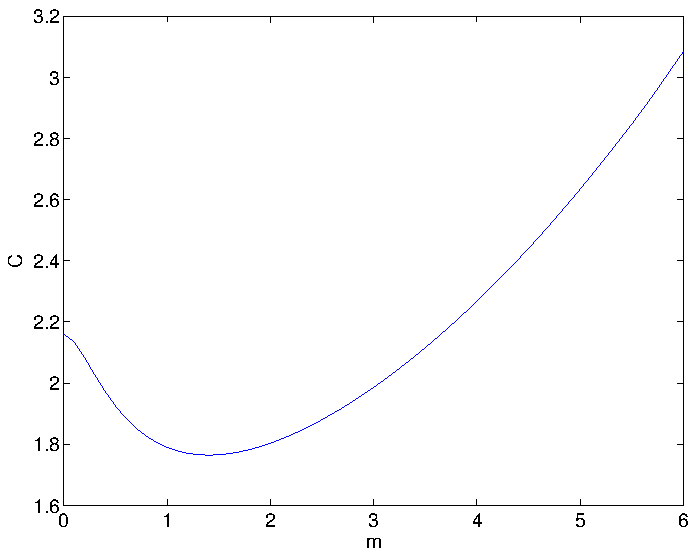}
\caption{The value of the constant $C_{m_\pi}$ as a function of the pion mass $m_\pi.$ \label{Cm}}
\end{center}
\end{figure}

We can generalise this approach in two ways. Firstly, we can introduce the pion mass $m_\pi \neq 0.$ Then the interaction energy can be written as 
\begin{equation}
\label{Eintma}
E_{{\rm int,} m_\pi}^{{\rm att}} = -\frac{2C_{m_\pi}^2}{3\pi}\exp(-m_\pi X)
\left(m_\pi^2 X^2+2m_\pi X+2\right)\frac{1}{X^3},
\end{equation}
in the attractive channel, \cite{Feist:2011aa}. Note that $C_{m_\pi}$ is now a function of the pion mass $m_\pi,$ which is plotted in figure \ref{Cm}. This figure agrees with the results in \cite{Feist:2011aa}.

A point worth noting is that we find $C_{m_\pi}=2.16$ for $m_\pi=0$ as in \cite{Manton,Feist:2011aa}. We also calculated $C_{m_\pi} = 1.93$ and $C_{m_\pi} = 1.79$ for $m_\pi=0.5$ and $m_\pi=1,$ respectively. These are the values of $m_\pi$ which will be important later.

As a second generalisation we also include relativistic corrections since we are interested in describing high velocities. The relativistic Lagrangian for point particles interacting via a radial potential $V$ is given by
\begin{equation}
\label{Lrel}
L_{{\rm point}}= -\sum\limits_{k=1}^{2} M\sqrt{1 - ({\bf v}^{(k)})^2} 
- V\left(|{\bf r}^{(1)} -{\bf r}^{(2)}|\right), 
\end{equation}
where ${\bf r}^{(k)}$ and ${\bf v}^{(k)} = \frac{d}{dt}{\bf r}^{(k)}$ are position and velocity of the $k$th particle.\footnote{Here we treat the particles relativistically, but we make the approximation that $V$ can be treated as a function of separation only -- ignoring retarded potentials.}
Note that the relativistic momentum is given by 
\begin{equation}
{\bf p}^{(k)} = \nabla_{{\bf v}^{(k)}} L_{{\rm point}} 
= \frac{M {\bf v}^{(k)}}{\sqrt{1 - ({\bf v}^{(k)})^2} }.
\end{equation}
The Euler-Lagrange equations then result in the usual force law
\begin{equation}
\label{Lag12eq}
\frac{d {\bf p}^{(k)}}{dt} = {\bf F}^{(k)}, \quad {\rm where} \quad
{\bf F}^{(k)} = -\nabla_{{\bf r}^{(k)}} V\left(|{\bf r}^{(1)} -{\bf r}^{(2)}|\right).
\end{equation}
In the following, we work in the centre of momentum frame ${\bf p}^{(1)} = - {\bf p}^{(2)},$  and we restrict our consideration to the nontrivial part of the attractive channel, namely, ${\bf r}^{(1)}=-{\bf r}^{(2)},$ with ${\bf r}^{(1)} \cdot {\bf n} =0$ and ${\bf p}^{(1)}\cdot {\bf n} =0.$\footnote{There are different definition of a relativistic centre of mass in the literature. Working in the centre of momentum frame avoids these difficulties.} Then we can use the identity
\begin{equation}
\nabla_{{\bf r}^{(1)}} V\left(|{\bf r}^{(1)} -{\bf r}^{(2)}|\right)
= -\nabla_{{\bf r}^{(2)}} V\left(|{\bf r}^{(1)} -{\bf r}^{(2)}|\right)
\end{equation}
to show that if (\ref{Lag12eq}) is satisfied for $k=1$ it is also satisfied for $k=2.$ The relativistic particle equations of motion become
\begin{equation}
\label{eomrel}
\frac{d^2{\bf r}}{dt^2} = \frac{1}{M \gamma({\bf v})} \left({\bf F} - \frac{({\bf F}\cdot {\bf v}) {\bf v}}{c^2}\right),
\end{equation}
where, for simplicity, we have suppressed the superscripts and 
\begin{equation}
{\bf F} = -\frac{{\bf r}}{|{\bf r}|} \left. 
\frac{dV(R)}{d R}\right|_{R=|2{\bf r}|}.
\end{equation}
The relativistic particle equations of motion (\ref{eomrel}) can now be solved for the dipole approximation $V(X)=E_{{\rm int}}^{{\rm att}}(X)$ in (\ref{Eint0a}), or the interaction potential for massive pions $V(X)= E_{{\rm int,} m_\pi}^{{\rm att}}(X)$ in (\ref{Eintma}). Since we are interested in scattering processes our initial conditions are that the Skyrmions are located at $\pm \frac{1}{2}(D,b,0)$ with initial velocities $\mp \frac{1}{2}(v,0,0)^T.$ This gives the initial conditions for ${\bf X}$ as ${\bf X}(0) = (D,b,0)^T,$ and ${\dot {\bf X}} = -(v,0,0)^T.$ Hence the relative angular momentum (\ref{lrel}) is ${\bf l}_{{\rm rel}} = \mu (0,0,bv)^T.$  Scattering is defined in the limit $D \to \infty.$ For finite $D$ not all velocities $v$ correspond to scattering solutions. For example for $m_\pi=0$ in the dipole approximation, starting with $b=0$ and $v=0$ at infinity ($D=\infty$) gives rise to $v=0.08$ at $D=10$ by energy conservation. If $v$ is chosen lower than $0.08$ at $D=10$ then the trajectories cannot escape to infinity.

\begin{figure}[!ht]
\begin{center}
\includegraphics[width=7.1cm]{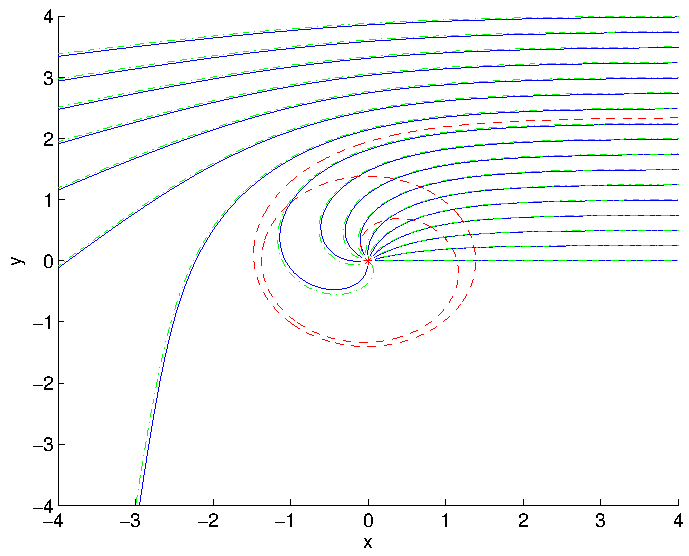}
\includegraphics[width=7.1cm]{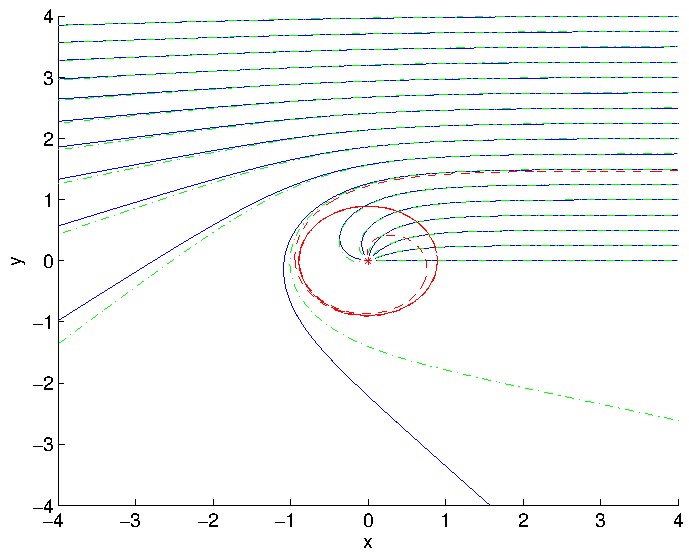}
\caption{Scattering trajectories of two Skyrmions in the dipole approximation with zero pion mass $m_\pi=0.$ The solid blue lines are the position of one Skyrmion for impact parameters $b=0,0.5,\dots,8.$ The dashed red line is the trajectory for the critical value of the impact parameter $b.$ The dashed-dotted green lines are the non-relativistic approximation. In the left figure the initial speed $v$ of one Skyrmion is $v=0.2$ with $b_{{\rm crit}}/2 = 2.35,$ while in the right figure $v=0.4$ and $b_{{\rm crit}}/2 = 1.47.$
\label{Fig2}}
\end{center}
\end{figure}

In figures \ref{Fig2} we show the trajectories in the dipole approximation $(m_\pi=0)$ with $v = 0.2$ and $v=0.4$ for various $b.$ Here $D$ is chosen sufficiently large. As can be seen from figure \ref{Fig2}, the Skyrmions attract each other for small impact parameter $b$ and collide at the origin when the particle equation of motion is no longer well defined. Once the impact parameter $b$ is larger than a critical value $b_{{\rm crit}}$ we observe scattering behaviour. The collision at the origin is an artefact of our approximation which does not include any short range repulsive force. Therefore the trajectories are only physical for $b > b_{{\rm crit}}.$ We plot both relativistic and non-relativistic dynamics and can see that there is reasonable agreement for most trajectories. The error becomes particularly noticeable for trajectories close to the critical impact parameter $b_{{\rm crit}}.$
\begin{figure}[!ht]
\begin{center}
\includegraphics[width=10cm]{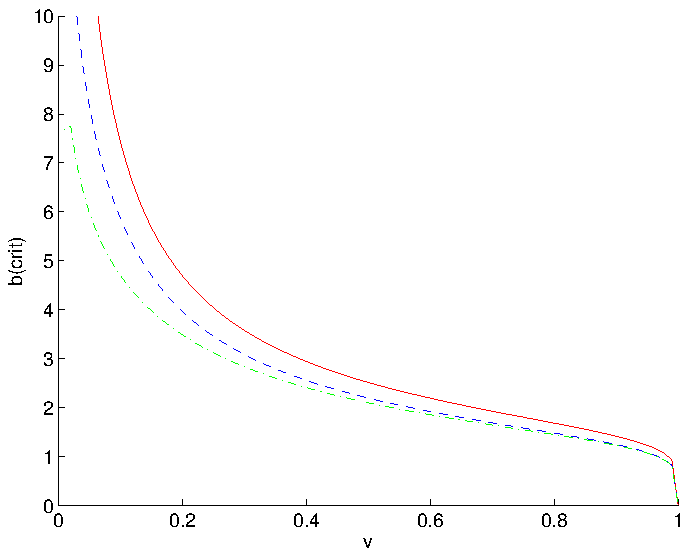}
\caption{The critical impact parameter $b_{{\rm crit}}$ as a function of the initial velocity $v.$ The solid red line corresponds to $m_\pi=0,$ the dashed blue line to $m_\pi = 0.5$ and the dashed-dotted green line to $m_\pi=1.$\label{Fig3}}
\end{center}
\end{figure}
Figure \ref{Fig3} shows the critical impact parameter $b_{{\rm crit}}$ as a function of $v$ for $m_\pi= 0.$ As can be expected $b_{{\rm crit}}$ decreases as $v$ increases, and $b_{{\rm crit}}$ tends to zero in the limit $v \to 1.$ In figure \ref{Fig4} we show how the scattering changes when the pion mass is increased to $m_\pi = 0.5.$ The scattering becomes less pronounced and $b_{{\rm crit}}$ is smaller than in the massless case. Figure \ref{Fig3} also shows the critical impact parameter $b_{{\rm crit}}$ as a function of $v$ for $m_\pi=0.5$ and $m_\pi = 1.$

\begin{figure}[!ht]
\begin{center}
\includegraphics[width=7.1cm]{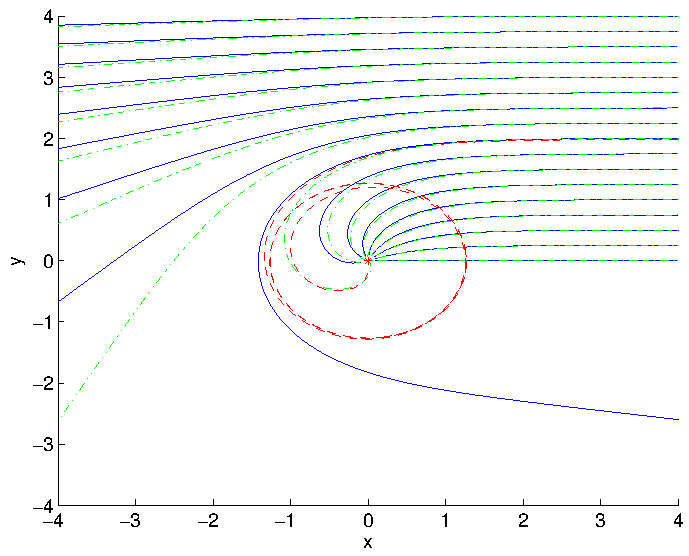}
\includegraphics[width=7.1cm]{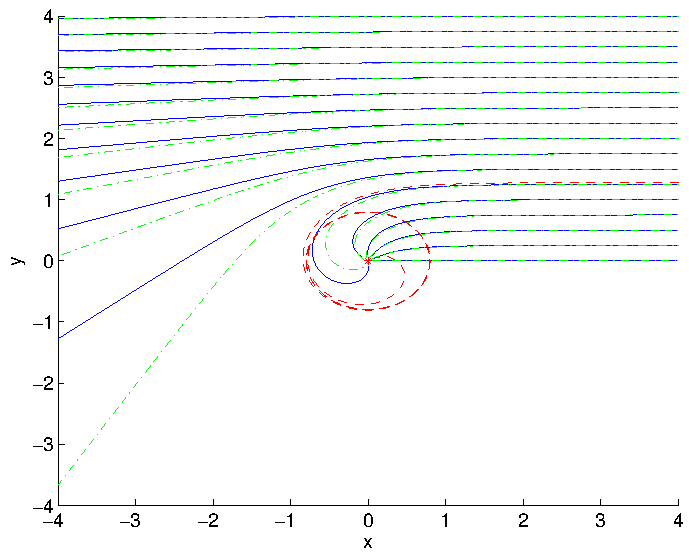}
\caption{Scattering trajectories of two Skyrmions in the dipole approximation with pion mass $m_\pi=0.5$ The solid blue lines are the position of one Skyrmion for impact parameters $b=0,0.5,\dots,8.$ The dashed red line is the trajectory for the critical value of the impact parameter $b.$ The dashed-dotted green lines are the non-relativistic approximation. In the left figure the initial speed $v$ of one Skyrmion is $v=0.2$ with $b_{{\rm crit}}/2 = 1.99,$ while in the right figure $v=0.4$ and $b_{{\rm crit}}/2 = 1.28.$
\label{Fig4}}
\end{center}
\end{figure}

Tracking the \sks location show how the pion mass, $m_\pi$, affects the scattering process. For example in figure \ref{Trajs}, it can be seen that, for large separation, the \sk with $m_\pi=0.5$ is deflected less. \Fig \ref{m0pt5-v0pt2-trajectories-one-side} also shows the two distinct scattering processes. One scattering process is for small $b$ where the \sks combine and then repel. This repulsion is a consequence of the geometry of the \sk moduli space and is analogous to monopole scattering. Only the scattering regime for large $b$ where the \sks are deflected towards each other can be approximated by dipole scattering. In fact, for $m_\pi=0$ in figure \ref{m0-v0pt2-trajectories-one-side} all scattering trajectories have $b<b_{\rm{crit}}$ and, therefore, cannot be described by the dipole approximation. For $m_\pi=0.5,$ the critical impact parameter is $b_{{\rm crit}}/2 = 1.47.$ Therefore, the outer three trajectories satisfy $b > b_{{\rm crit}}.$ The outer two show a qualitatively similar behaviour to the trajectories in figure \ref{Fig2}b while the third trajectories clearly experiences an additional repulsive force. 
For intermediate impact parameters, $b\sim 0.6,$  the geometric effect and the dipole attraction compete. A point worth noting is that this competitive effect might effectively cancel for a scattering with impact parameter between $0.6$ and $0.8$, for both $m_\pi=0$ and $m_\pi=0.5$. This is the range of trajectories for which the $y$ value at $x=1.5$ swaps from being below the corresponding impact parameter to above it. This cancelation would not give a flat trajectory, but it would have the same $x$ values at $y=1.5$ as at $y=-1.5$. \Fig \ref{m0-v0pt2-trajectories-one-side} shows that the trajectories for $b=1.8$ and $b=2$ crossover for $m_\pi=0$. It can be seen from \fig\ref{m0pt5-v0pt2-trajectories-one-side} that does not happen for the same trajectories when $m_\pi=0.5$. This is can be understood because  both the geometric repulsive and dipole attraction effects are less for the increasingly localised $m_\pi=0.5$ \sk. In our discussion, we have assumed that the Skyrmions are not spinning. 
When Skyrmions spin further $\frac{1}{r}$ terms from the kinetic
term in the full Lagrangian \eqref{Lag} contribute to the point
particle Lagrangian \eqref{Lrel}.
These contribution have been independently calculated by Schroers \cite{Schroers:1993yk} and Gisiger and Paranjape \cite{Gisiger:1994gj, Gisiger:1994pr, Gisiger:1998tv}. As these terms are of order $\frac{1}{r}$ they could dominate over the dipole interaction and could lead to profoundly different scattering trajectories.

\section{Skyrmions visualisation}
\label{Visualisation}

For a long time Skyrmions have been visualised as level sets of baryon density, and recently it has become standard practice to colour the level sets in order to show the value of the pion fields. This is a good method to visualise \sks, especially as it uses an invariant of the model. It clearly displays the symmetry of \sks and shows how the \sks can  potentially be combined to make larger \sks. It is also a good method to visualise \sk scattering as shown in the previous images. But level sets of baryon density do not show how the \sks recombine during a scattering process. For example, as previously shown, when two \sks are in the attractive channel and collide head-on then they scatter perpendicularly. From the simulations it seems as though half of each \sk is exchanged, and the corresponding two halves recombine to make two new \sks travelling perpendicularly to the original velocities. Our aim is to quantify this exchange and to visualise it in a new way which could shine light onto \sk dynamics. 
Our construction is to track the preimages, $U(p_i)^{-1}$, of a range of points $p_i \in SU(2)$ throughout a collision. Note that $U(p)^{-1}$ denotes the set of preimages of the point $p$ and is not to be confused with the matrix inverse which is given by $U^\dagger$ for unitary matrices $U.$

\subsection{Preimages} \label{preimages}

So far we have defined the location of \sks as the points $U(-1_2)^{-1}$ in $\mathbb{R}^3$. We shall now describe how we chose the preimages to track. 

Our aim is to visualise a \sk scattering using preimages. Our initial configuration and initial velocities are symmetric under the combined reflections 
\begin{equation}
\left(
\begin{array}{r} 
x\\ y\\ z
\end{array} \right) 
\mapsto
\left(
\begin{array}{r} 
x\\ y\\ -z
\end{array} \right)
\quad {\rm and} \quad 
\left(
\begin{array}{r} 
\pi_1\\ \pi_2\\ \pi_3
\end{array} \right) 
\mapsto
\left(
\begin{array}{r} 
\pi_1\\ \pi_2\\ -\pi_3
\end{array} \right), 
\end{equation}
where $z=0$ corresponds to the scattering plane. The reflection symmetry implies that $\pi_3=0$ in the scattering plane, namely $\pi_3(x,y,0)=0.$  Hence we can define the equatorial two-sphere as $S^2_{{\rm eq}} =\{(\sigma, \pi_i)|\sigma^2+\pi_1^2+\pi_2^2 = 1, \pi_3=0\} \subset S^3 \cong SU(2)$. Then, for a single $B=1$ hedgehog \eqref{hog} all of the points $U(p_i)^{-1}$ in $\mathbb{R}^3$ for $p_i\in S^2_{{\rm eq}}$ will lie on the scattering plane. This gives a two-dimensional way to visualise the three-dimensional \sk using preimages which lie in the scattering plane, namely,  we track the preimages of points in $S^2_{{\rm eq}}$ to visualise a scattering process. 

As much as we would like to, numerically we cannot track all of the preimages of $S^2_{{\rm eq}}$. As we know, \sks in this model are not discrete objects, but they are actually extended objects. When visualising a two-\sk solution, with large separation, as a level set of \bdn we have to arbitrarily choose a value of \bdn which shows two distinct \sks. As our aim is to use preimages to represent a two-\sk system, where we can identify single \sks, we choose a cut-off and do not sample points on $S^2_{{\rm eq}}$ where $\sigma >0.5.$ This is an arbitrary aesthetic choice. A cut-off is needed, so that we do not track points too near to the vacuum, $\sigma =1$. These points can move very rapidly due to radiation propagating around the system since perturbations about the vacuum have very little mass. Therefore, tracking points near the vacuum would give an unrealistic representation of the collision. 

\begin{figure}[H]
       \centering
       \begin{subfigure}[b]{0.4\textwidth}
               \centering
    \includegraphics[width=\textwidth]{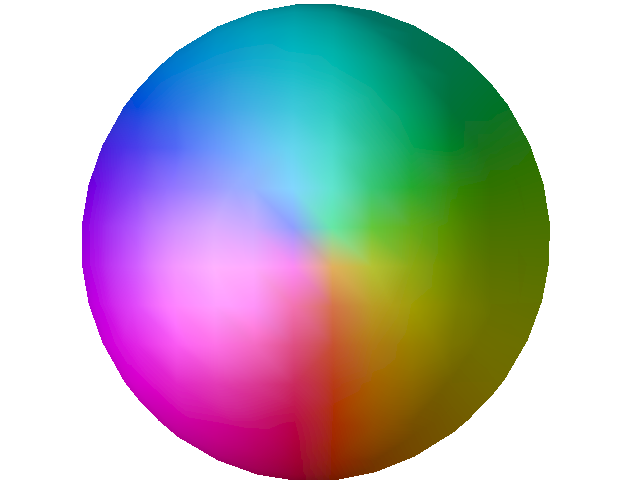}
               \caption{~}
               \label{Single-skyrmion-colored} 
       \end{subfigure}%
        ~ 
        \begin{subfigure}[b]{0.4\textwidth}
                \centering \includegraphics[width=\textwidth]{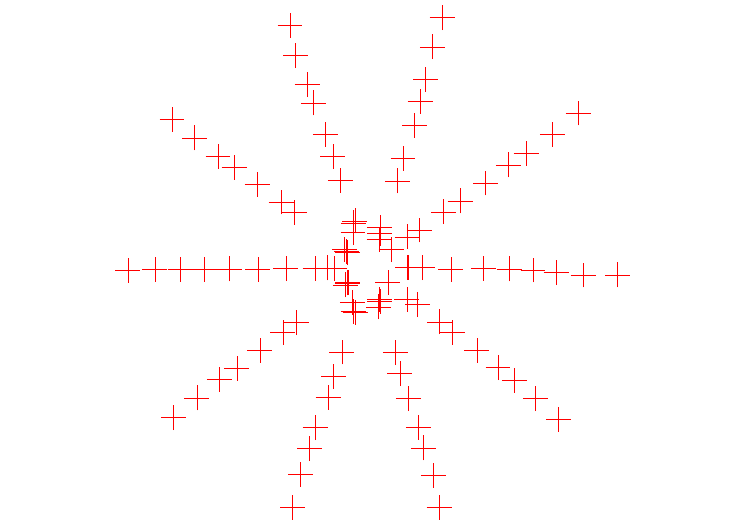}
                \caption{~}
                \label{single-palm}
        \end{subfigure}
        ~ 
        \caption{Comparison of the \bd plot, \ref{Single-skyrmion-colored}, and the preimage plot, \ref{single-palm}, of a single \sk.}
\end{figure}

We chose to track the points,
\begin{align}\label{points}
\sigma_k &=\frac{1}{2}-\frac{3k}{2 k_{\max}}, \\ \nonumber
\pi_{1,n}&=\sqrt{1-\sigma_k^2} \cos\left(\frac{2\pi n}{n_{\max}-1}\right), \\ \nonumber
\pi_{2,n}&=\sqrt{1-\sigma_k^2} \sin\left(\frac{2\pi n}{n_{\max}-1}\right), \nonumber
\end{align}
where $k$ and $n$ are integers, $1\leq k<k_{\max}$ and $1\leq n<n_{\max}$. This range is appropriate because if $k=k_{\max}$ then there would be $n_{\max}$ points where $\sigma=-1, \pi_1=\pi_2=0.$ Hence this value of $k$ is excluded. The preimage of $\sigma = -1$ is also the location which we have already tracked. This defines $(n_{\max}-1)(k_{\max}-1)$ points on $S^2_{{\rm eq}}$. \Fig \ref{single-palm} shows, for a single $B=1$ \sk, the preimages of $n_{\max}=k_{\max}=11$ points given by \eqref{points}, and compares it with the standard \bd level-set image in \fg \ref{Single-skyrmion-colored}. Note that for a two-\sk configuration there are $2(n_{\max}-1)(k_{\max}-1)$ points.

\begin{figure}[H]
       \centering
       \begin{subfigure}[b]{0.4\textwidth}
               \centering
    \includegraphics[width=\textwidth]{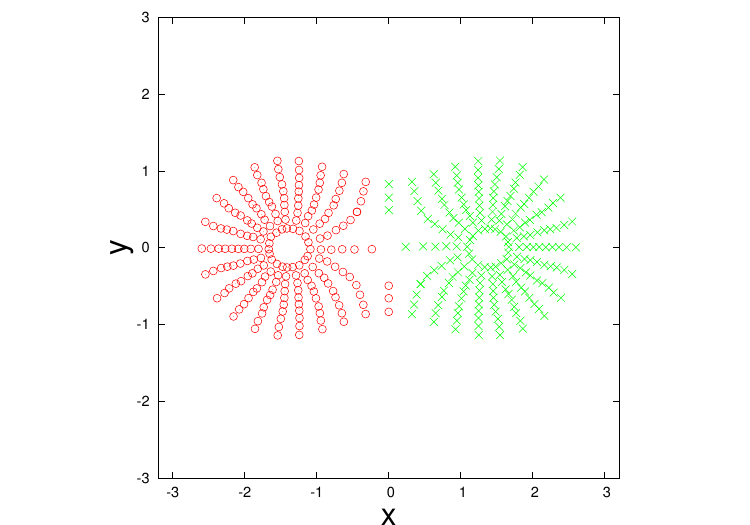}
               \caption{Before collision.}
               \label{t0perp} 
       \end{subfigure}%
        \! 
        \begin{subfigure}[b]{0.4\textwidth}
                \centering \includegraphics[width=\textwidth]{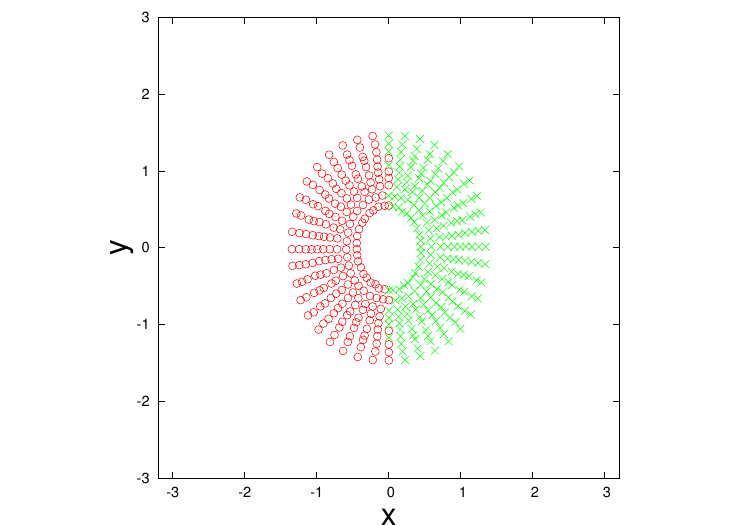}
                \caption{$B=2$ toroidal \sk.}
                \label{t1perp}
        \end{subfigure}
              \begin{subfigure}[!htb]{0.4\textwidth}
                \centering \includegraphics[width=\textwidth]{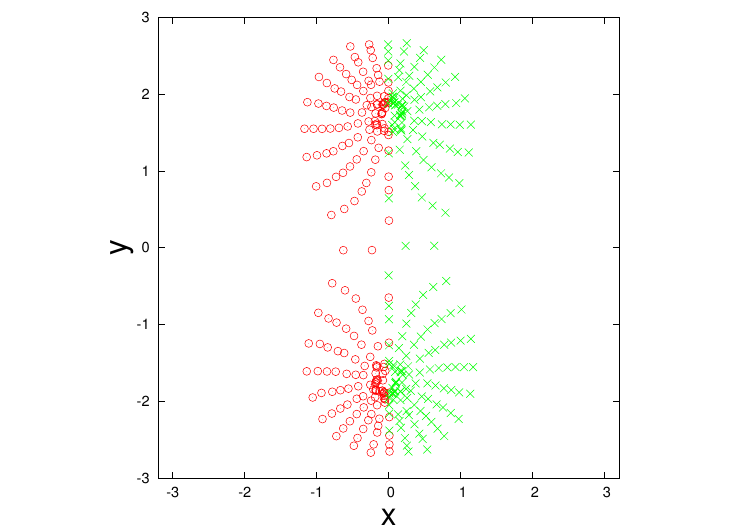}
                \caption{After collision.}
                \label{t2perp}
        \end{subfigure}
        \caption{Preimages of perpendicularly scattering \sks}
            \label{gap-0-palm}
\end{figure}

For each time slice we tracked the movement of each preimage using a search algorithm to find the point in $\mathbb{R}^3$ which has the required field value and is the closest to the same point of the previous time step. We are only interested in tracking how the preimages in the initial configuration move. It should be noted that the algorithm interpolated the field values in between the lattice sites to increase accuracy. This gives us a new insight into scattering. We can now see how the preimages move during a scattering process. For example, for $b=0$ the preimages scatter perpendicularly giving \fg \ref{gap-0-palm}. 

\begin{figure}[H] 
       \centering
       \begin{subfigure}[b]{0.4\textwidth}
               \centering
    \includegraphics[width=\textwidth]{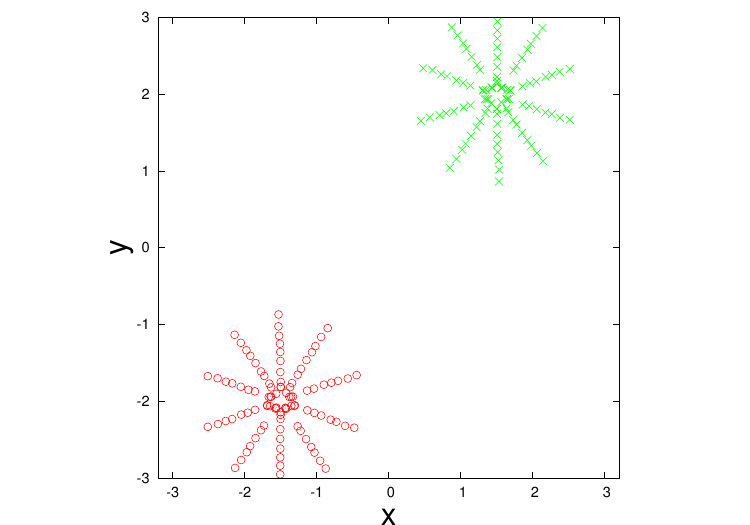}
               \caption{Before collision.}
               \label{gap-2-palm-scattering0} 
       \end{subfigure}%
        \! 
        \begin{subfigure}[b]{0.4\textwidth}
                \centering \includegraphics[width=\textwidth]{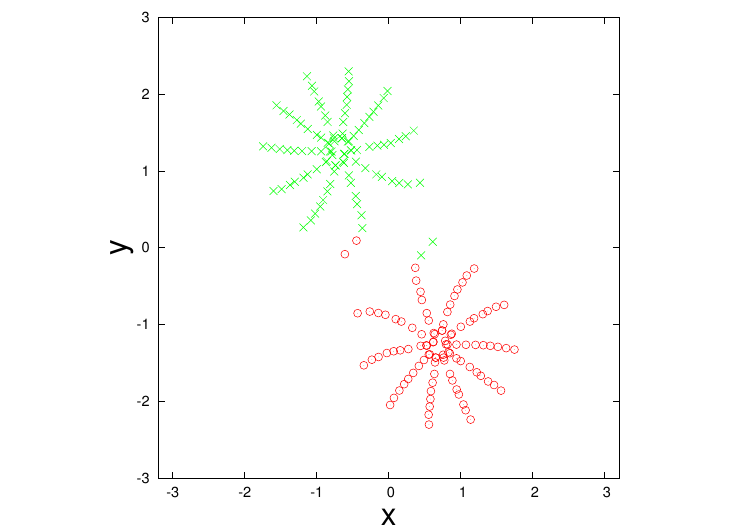}
                \caption{Preimage exchange}
                \label{gap-2-palm-scattering1}
        \end{subfigure}
              \begin{subfigure}[!htb]{0.4\textwidth}
                \centering \includegraphics[width=\textwidth]{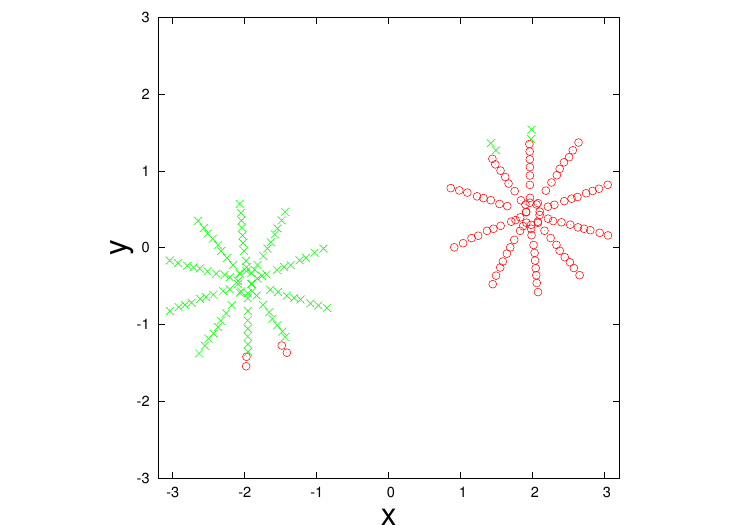}
                \caption{After collision.}
                \label{gap-2-palm-scattering2}
        \end{subfigure}
        \! 
        \begin{subfigure}[!htb]{0.4\textwidth}
                \centering \includegraphics[width=\textwidth]{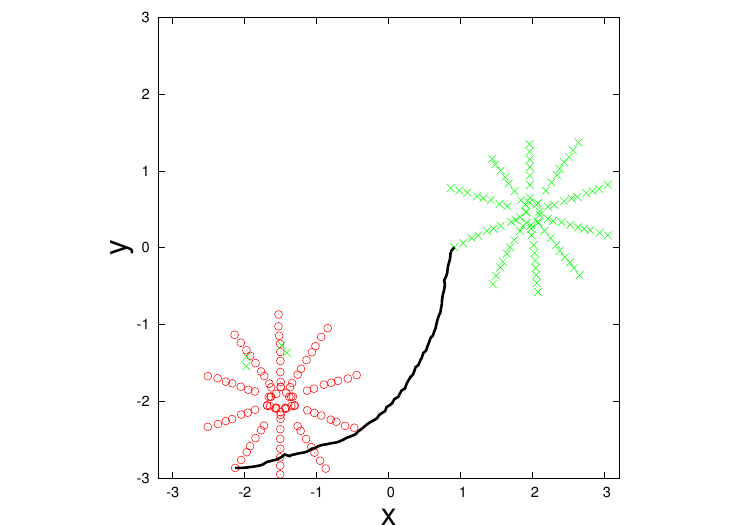}
                \caption{Initial and final preimages of the scattered \sk. Also shows the trajectory of one preimage explicitly showing the rotation.}
                \label{gap-2-palm-scattering3}
        \end{subfigure}
        \caption{Trajectories of the location of a single \sk throughout a scattering process ($k_{\max} =11, n_{\max}=11,m_\pi=0$).}\label{gap-2-palm}
\end{figure}

This new way of visualising \sks immediately shows that half of each \sk is used to form two new \sks, and the new recombined \sks are now rotated. This is the cause of the rotationless rotation observed previously. This is implied by the \bd plots, and it clearly shown in the preimage plots. What is not obvious from the \bd plots is that this preimage exchange also occurs for large impact parameters. An example of two \sks scattering with impact parameter $b=4$ is displayed in \fg \ref{gap-2-palm}. \Fig \ref{gap-2-palm-scattering0} shows the preimages of two \sks. In \fg \ref{gap-2-palm-scattering1} the two \sks exchange four preimages as they pass each other. \Fig \ref{gap-2-palm-scattering2} shows the preimages of the final scattered \sks. Hence, \sks do exchange preimages. Also, \fg \ref{gap-2-palm-scattering3} shows the initial preimages (red circles) and the final preimages (green crosses) of a single \sk. In \fg \ref{gap-2-palm-scattering3} we have also included the trajectory of a preimage. This shows that the \sk has rotated even for a large impact parameter.

\begin{figure}[H]
\centering
\includegraphics[width=10cm]{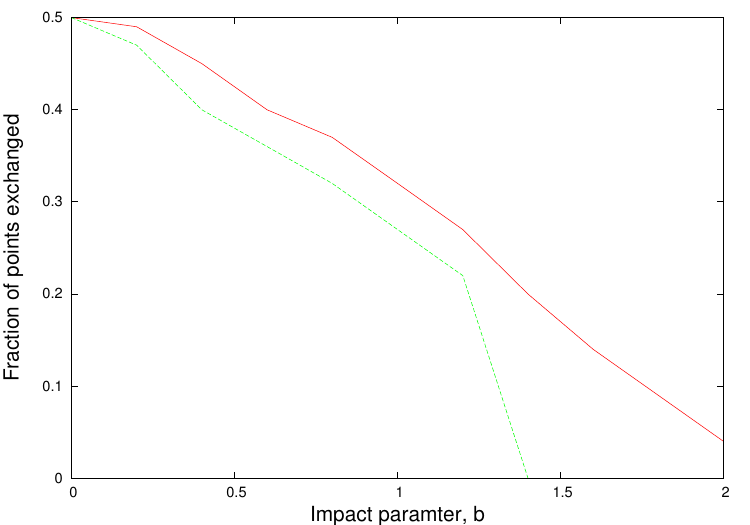}
\caption{Fraction of points exchanged for $m_\pi=0$ (solid line) and $m_\pi=1$ (dashed line).}
\label{points_ex}
\end{figure}

In our algorithm we were also able to track preimages for different scattering processes in order to quantify how many preimages are exchanged as a function of the impact parameter $b$. This is shown in \fg \ref{points_ex}. As the \sks pass each other they exchange preimages, and the number of exchanged preimages reduces with separation. This reduction in exchange is intuitive because \sks are localised objects.

\begin{figure}[H]
\centering
\includegraphics[width=10cm]{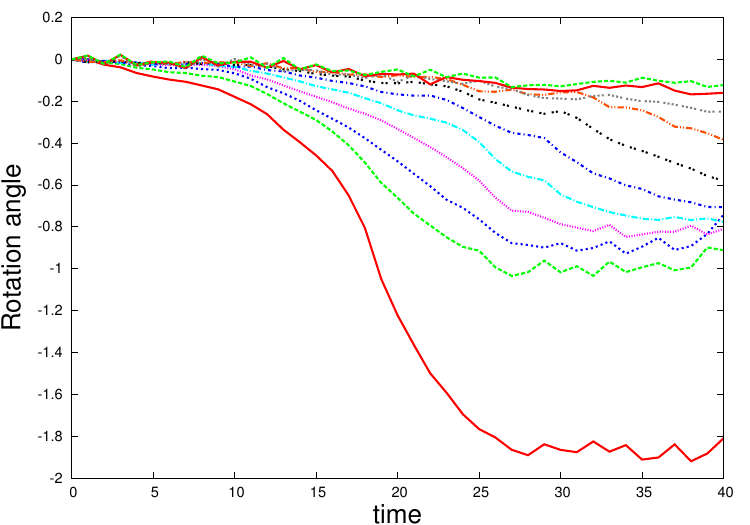}
\caption{Rotation of a single \sk throughout a scattering, $m_\pi=0$.}
\label{m0-rot}
\end{figure}

By carefully tracking these preimages we can measure the rotation angle of one \sk during a scattering process. We achieved this by tracking the relative orientation between the location point and the set of preimages which are constant $\pi_1,\pi_2$~-- this is one `arm' of the preimage plot in \fg \ref{single-palm}. Care must be taken not to choose points which are exchanged. By tracking the relative average orientation between the location and the set of points of constant $\pi_1,\pi_2,$ and not just one point, reduces the effect of radiation. The rotation angle is shown as a function of time in \fg \ref{m0-rot}. The oscillations in the rotation angle at large times are due to radiation propagating around the numerical lattice. \Fig \ref{m0-rot} shows that the \sks maximally rotate for $b=0$ when the rotation angle is approximately $\frac{\pi}{2}.$ The rotation angle decreases as $b$ increases. This is can be understood because \sks are localised objects. Hence as $b$ increases they exchange less preimages as they overlap less,  and therefore the \sks experience less rotation.

Another way of gaining an understanding of this phenomenon is to consider the attractive channel approximation in \cite{Schroers:1993yk}. Since the initial configurations are not spinning or isospinning, the total isospin ($M_2$ in \cite{Schroers:1993yk}) is zero. Since the total isospin is conserved, this sets the rotation angular frequency $\omega_2$ equal to the isorotation angular frequency $\Omega_2$ using the attractive Lagrangian in \cite{Schroers:1993yk}. Since both rotation and isorotation angles are zero, initially, they remain equal during the scattering process. If there was right-angle scattering, then the position of one Skyrmion would be rotated by $\frac{\pi}{2}$ and the phase would also be rotated by $\frac{\pi}{2},$ as observed in \fig \ref{m0-rot}. However, in this approximation, head-on collision does not lead to right angle scattering as the approximation breaks down for small separation. 

\subsection{Spinning \sks}
 Instead of simply colliding \sks, we also investigated colliding spinning \sks. We achieved this by numerically evolving an initial condition of two rotating hedgehog \sks boosted towards each other. We chose the \sks to be orientated in the attractive channel, and rotate in the same direction and angular frequency. This is similar to a constant global isorotation, and the Skyrmions remain in the attractive channel. It is known that for $m_\pi=0$ spinning \sks are not stable as they radiate pions \cite{Schroers:1993yk, Battye:2005nx}.  This is not a problem when we considered $m_\pi=0$  as the scattering takes place well before the \sks stop spinning. 

There has been some recent interest in spinning \sks, namely \cite{Hata:2010zy} and \cite{Hata:2011zz}, which investigate an extension of the collective coordinate quantisation procedure. The related question of isospin was examined in \cite{Battye:2014qva} where the authors considered the deformation introduced by isospinning \sks.

\begin{figure}[H] 
       \centering
 \begin{subfigure}[b]{0.3\textwidth}
               \centering
    \includegraphics[width=\textwidth]{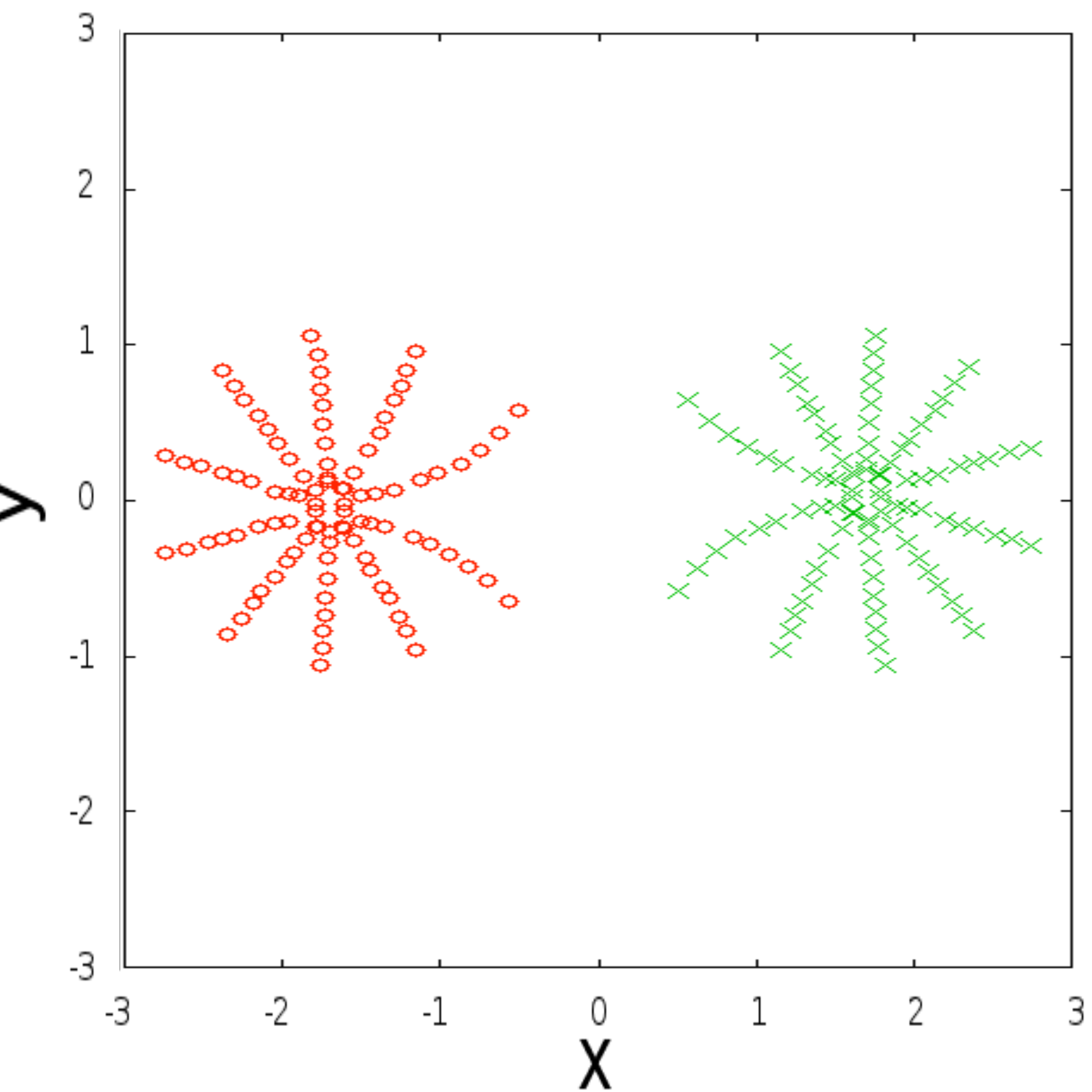}
              \caption{Before collision.}
               \label{palm-spinning1-s} 
       \end{subfigure}%
        \! 
\begin{subfigure}[b]{0.3\textwidth}
                \centering \includegraphics[width=\textwidth]{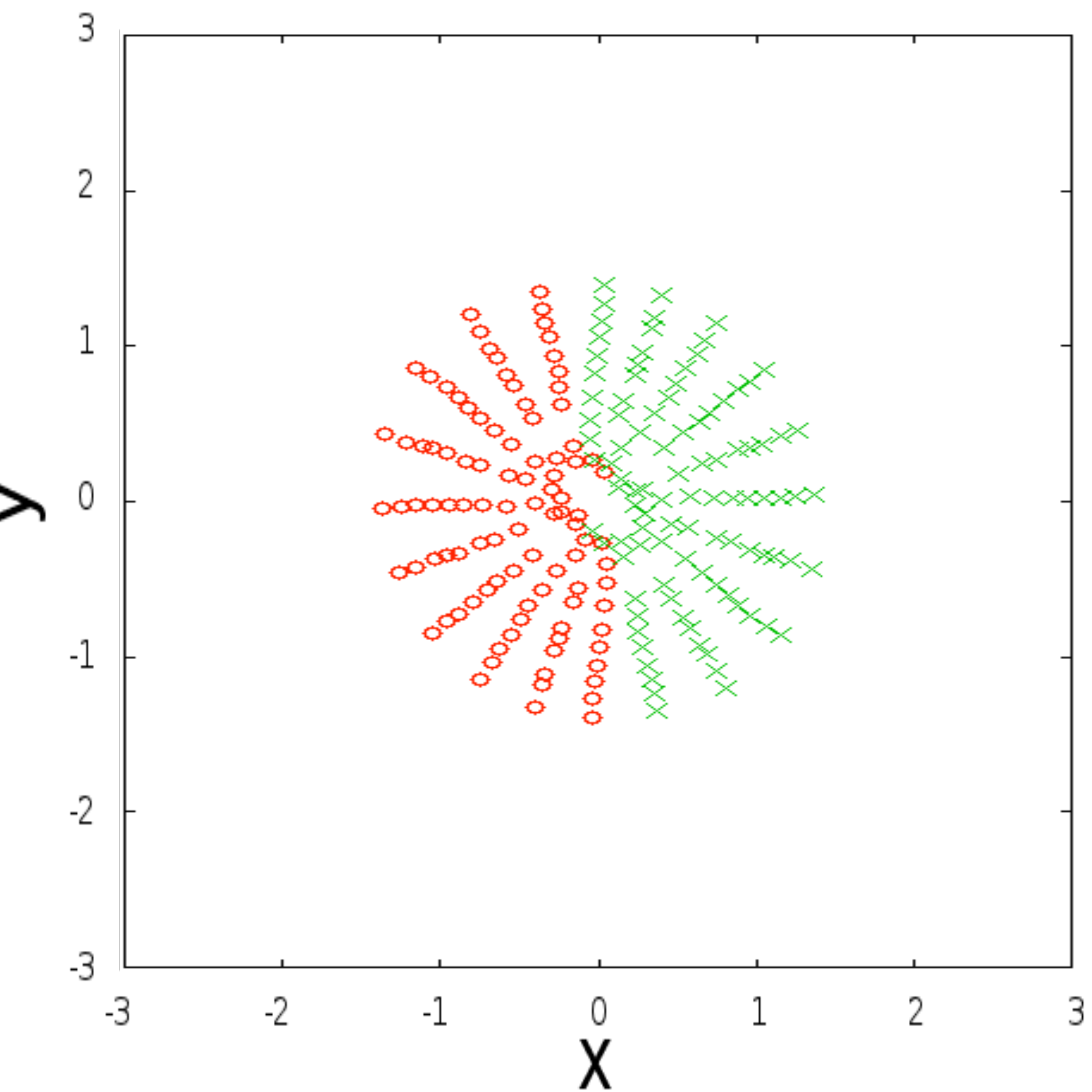}
                \caption{Preimage exchange}
                \label{palm-spinning2-s}
        \end{subfigure}
\begin{subfigure}[b]{0.3\textwidth}
                \centering \includegraphics[width=\textwidth]{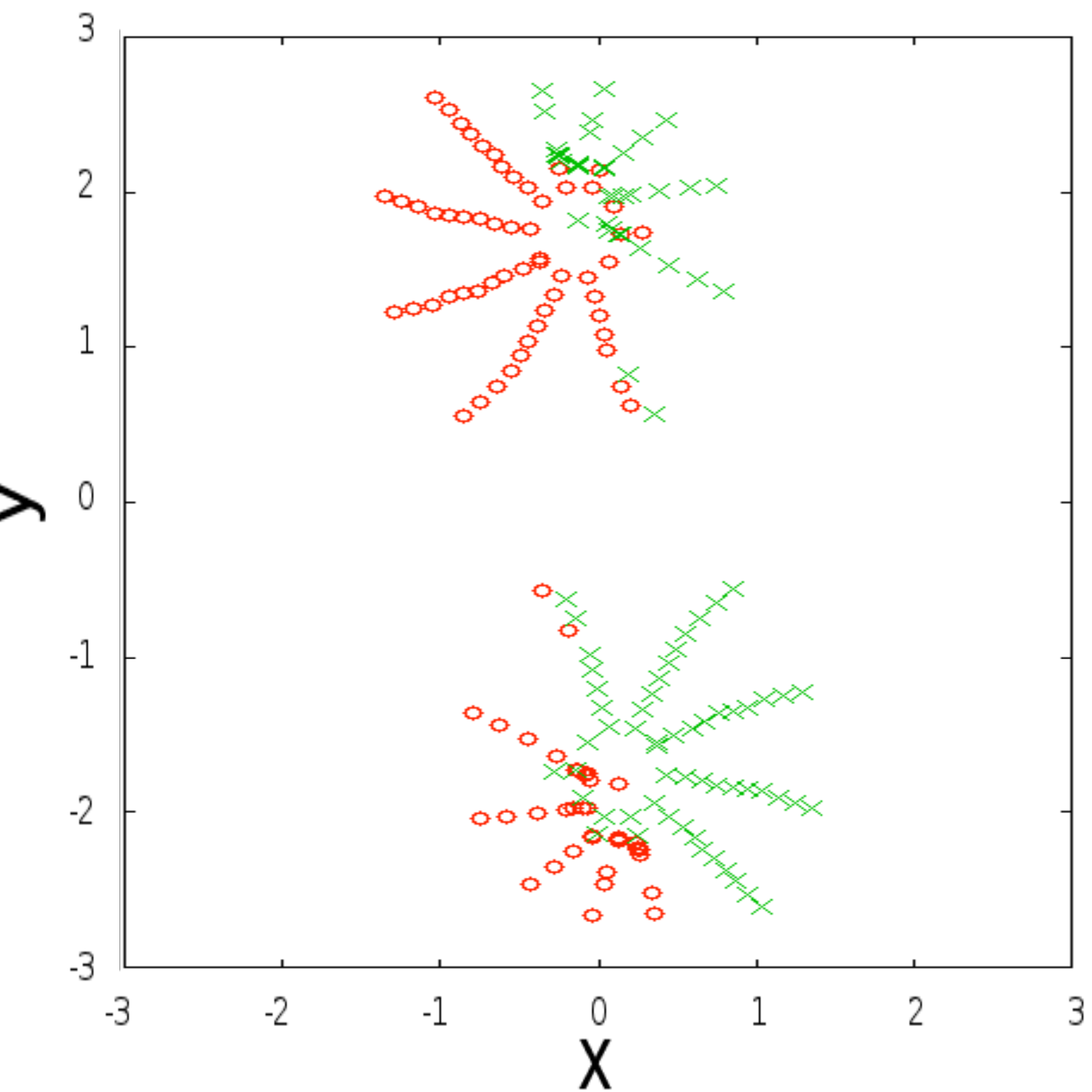}
                \caption{After collision.}
                \label{palm-spinning3-s}
        \end{subfigure}
        \caption{Preimage plots of two scattering \sks initially spinning at $0.05$ radians per unit time.}
        \label{palm-spinning-s}
\end{figure}

\begin{figure}[H] 
       \centering
 \begin{subfigure}[b]{0.3\textwidth}
              \centering
    \includegraphics[width=\textwidth]{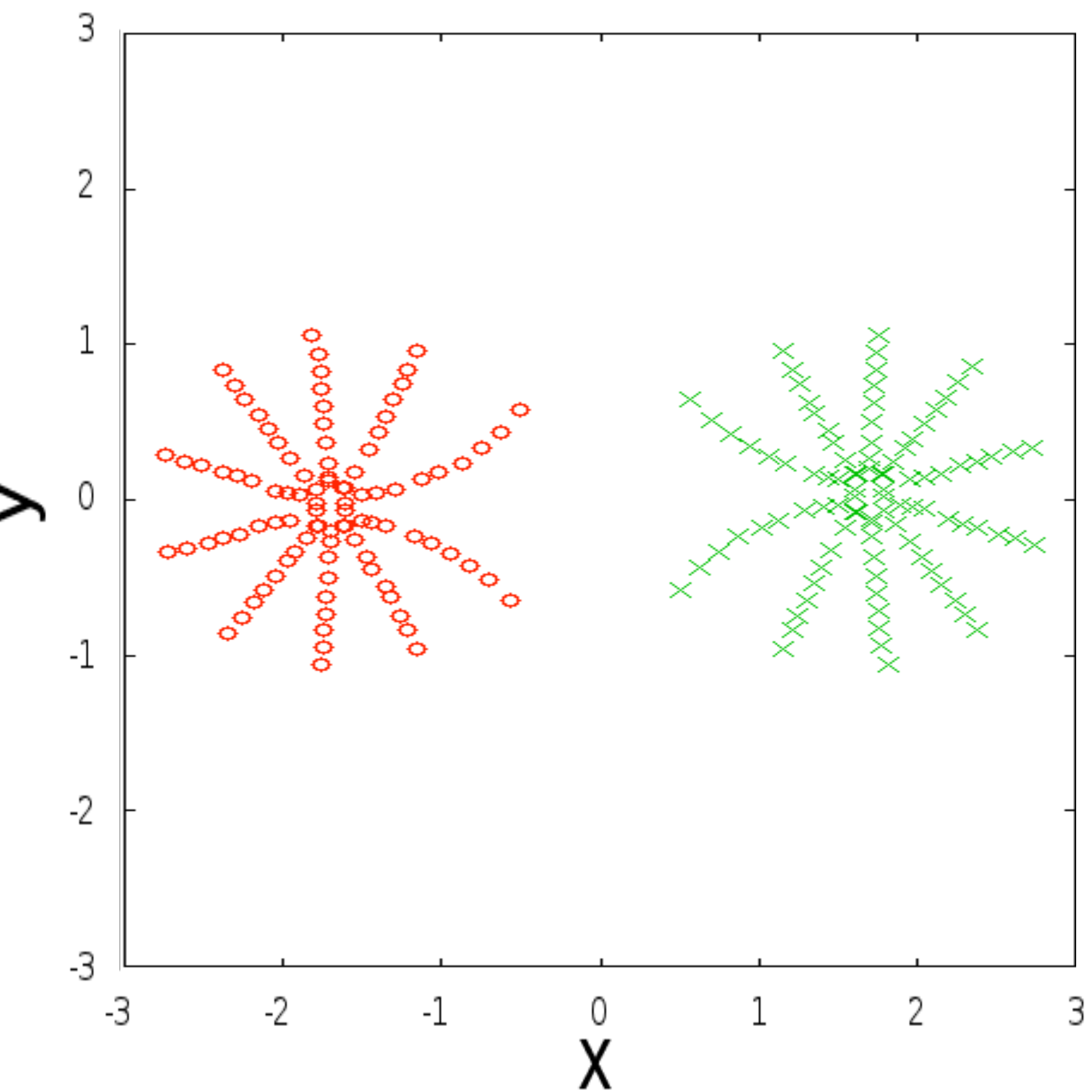}
               \caption{Before collision.}
               \label{palm-spinning1-m} 
       \end{subfigure}%
        \! 
 \begin{subfigure}[b]{0.3\textwidth}
                \centering \includegraphics[width=\textwidth]{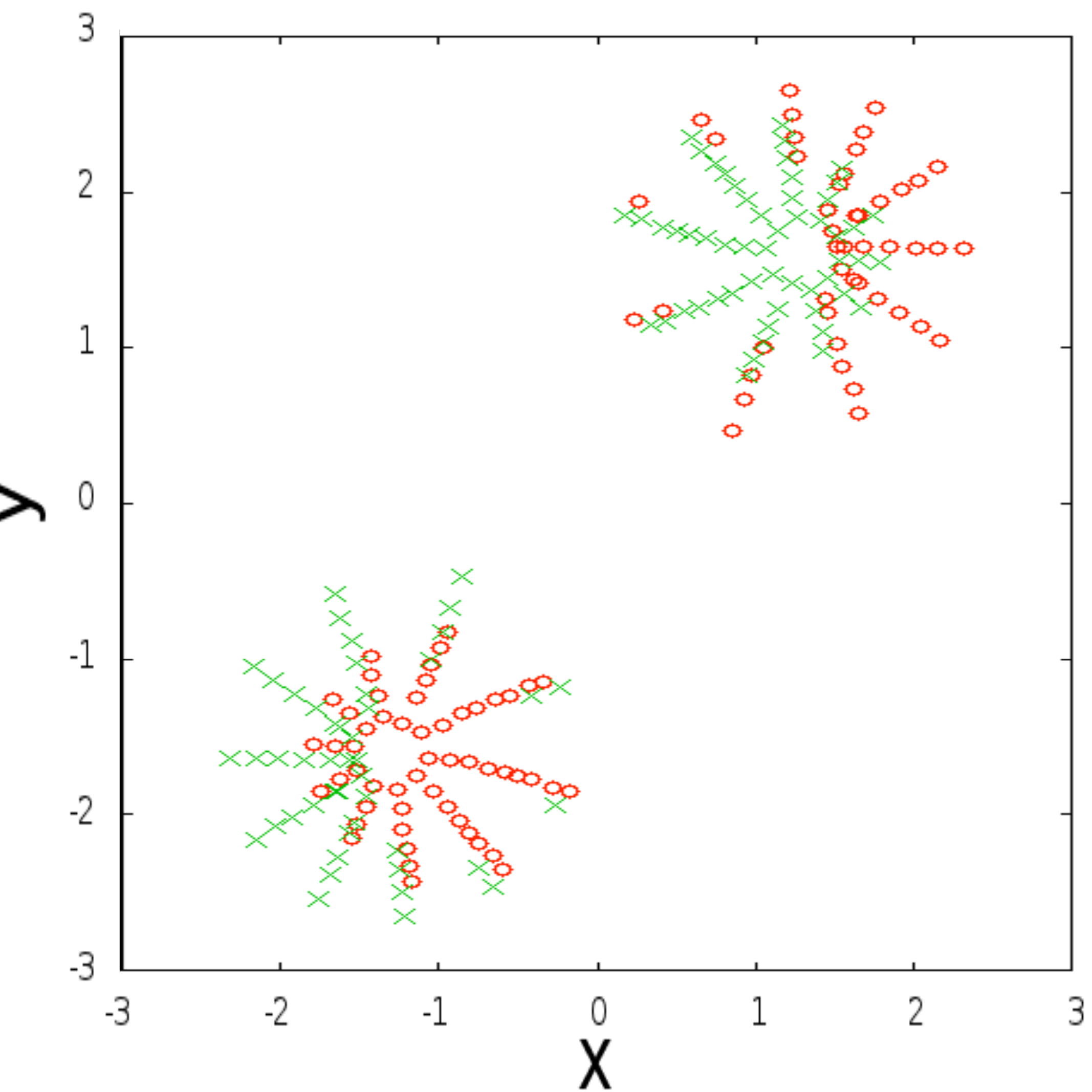}
                \caption{Preimage exchange}
                \label{palm-spinning2-m}
        \end{subfigure}
\begin{subfigure}[b]{0.3\textwidth}
                \centering \includegraphics[width=\textwidth]{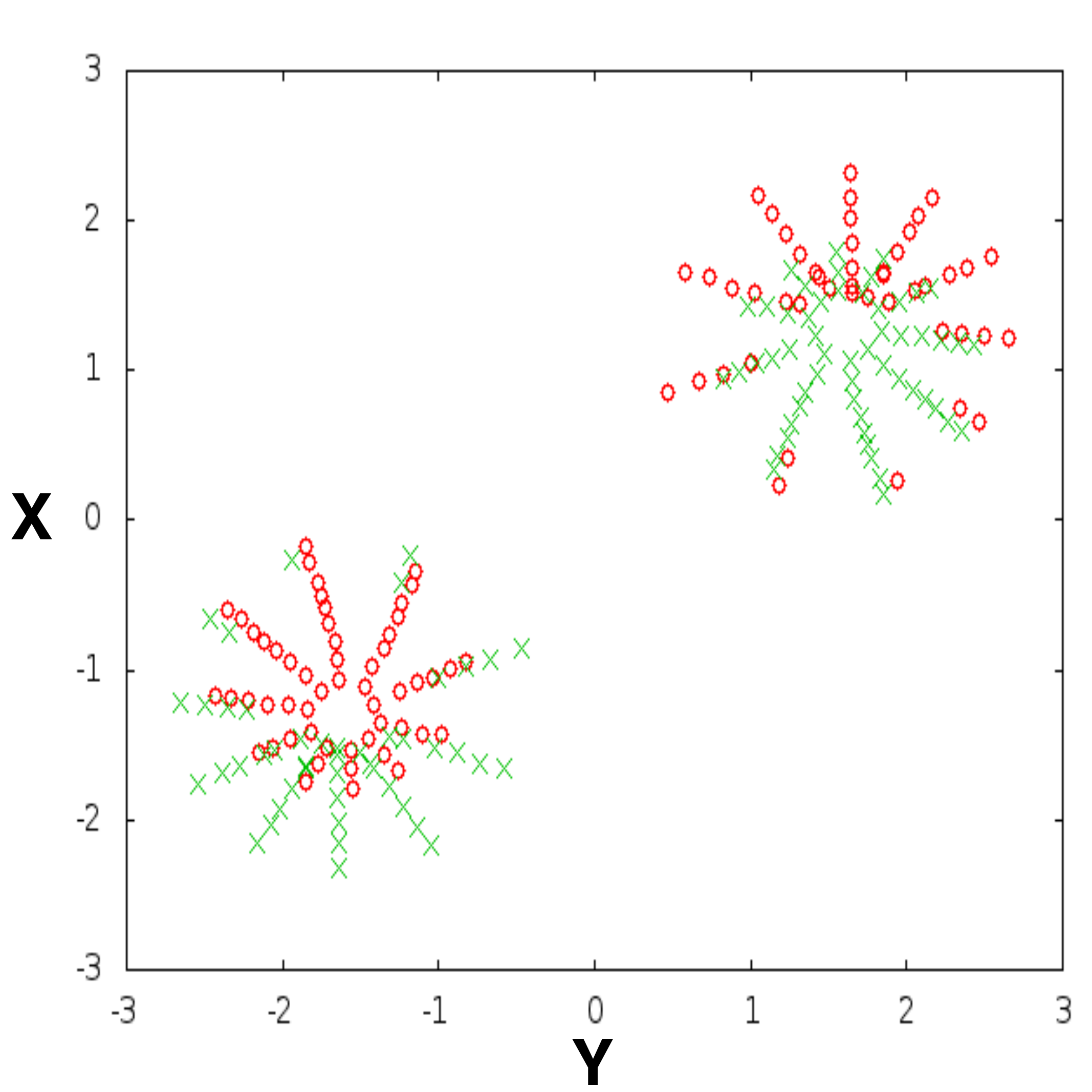}
               \caption{After collision.}
                \label{palm-spinning3-m}
        \end{subfigure}
        \caption{Preimage plots of two scattering \sks initially spinning at $0.5$ radians per unit time. }
        \label{palm-spinning-m}
\end{figure}

\begin{figure}[H] 
       \centering
\begin{subfigure}[b]{0.3\textwidth}
               \centering
    \includegraphics[width=\textwidth]{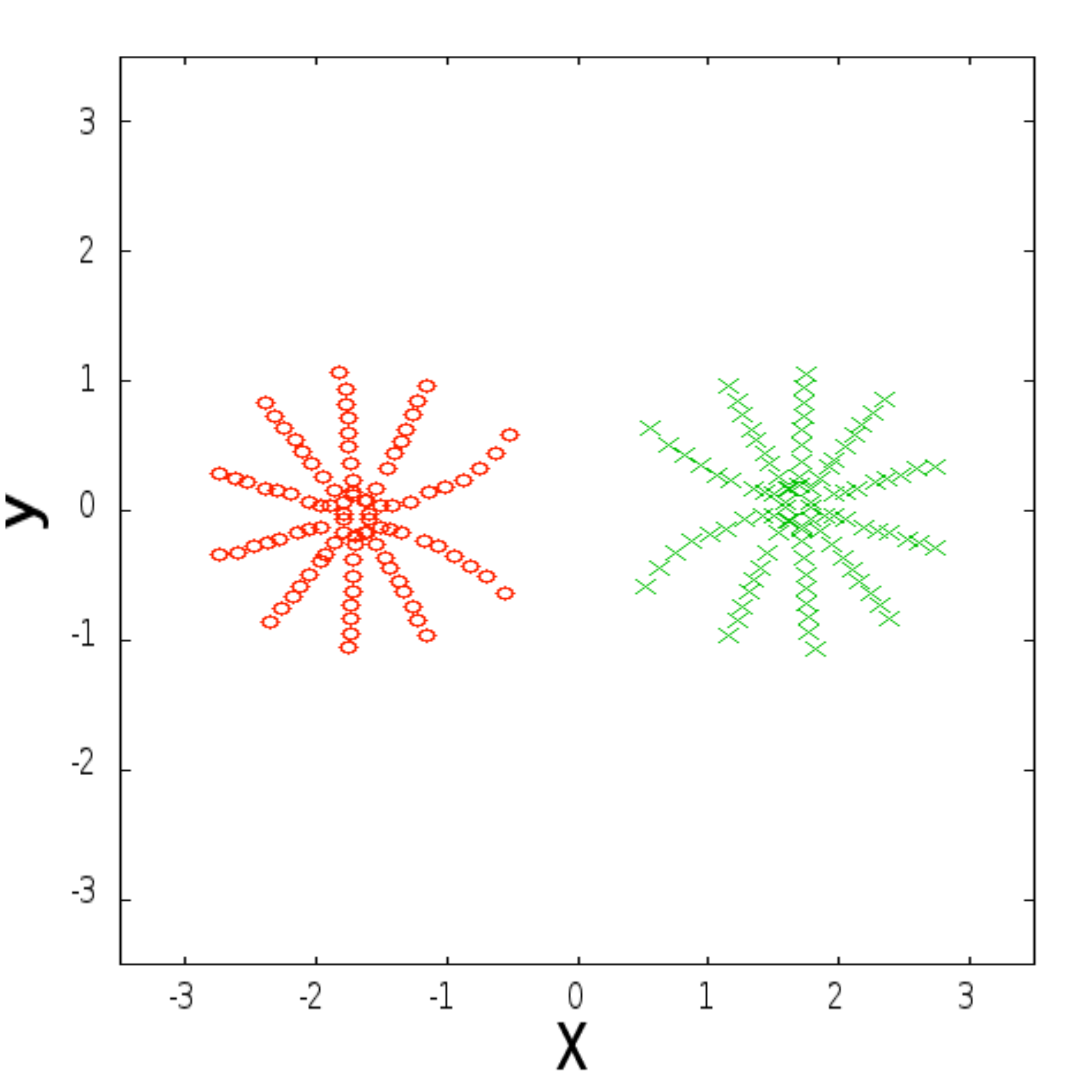}
               \caption{Before collision.}
               \label{palm-spinning1-f} 
       \end{subfigure}%
        \! 
\begin{subfigure}[b]{0.3\textwidth}
                \centering \includegraphics[width=\textwidth]{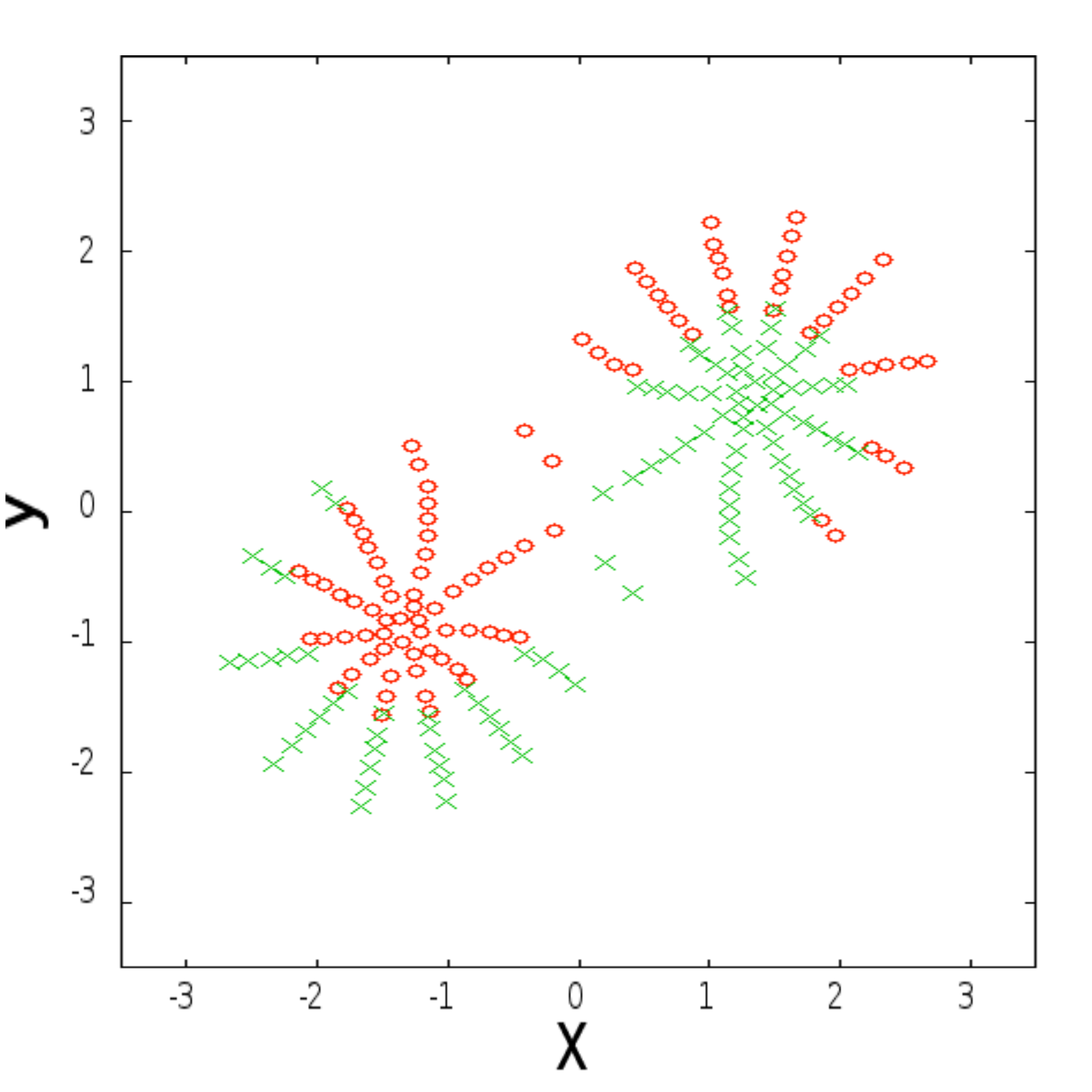}
                \caption{Preimage exchange}
                \label{palm-spinning2-f}
        \end{subfigure}
\begin{subfigure}[b]{0.3\textwidth}
                \centering \includegraphics[width=\textwidth]{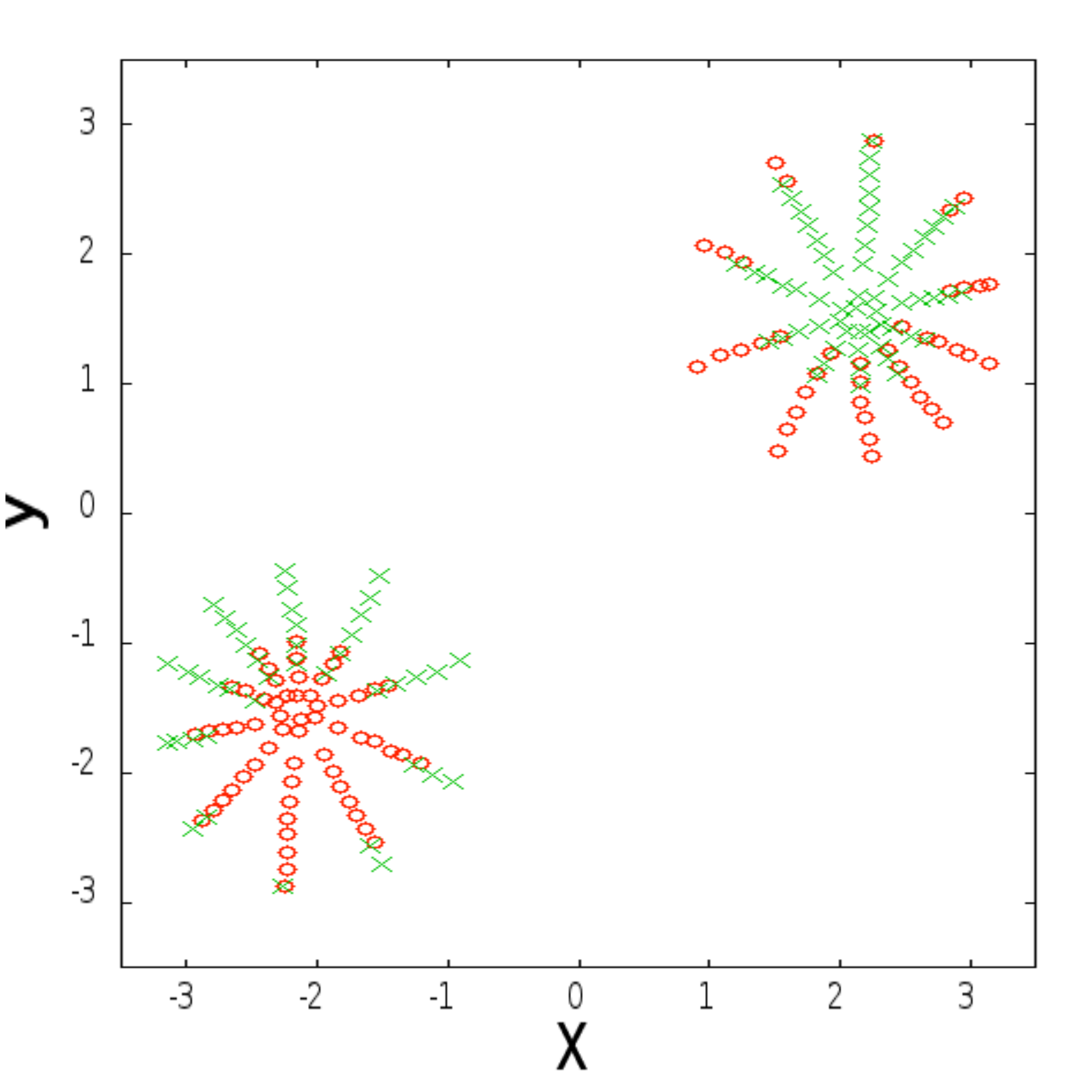}
               \caption{After collision.}
                \label{palm-spinning3-f}
        \end{subfigure}
        \caption{Preimage plots of two scattering \sks initially spinning at $1$ radians per unit time. }
        \label{palm-spinning-f}
\end{figure}

\begin{figure}[H]
\centering
\includegraphics[width=10cm]{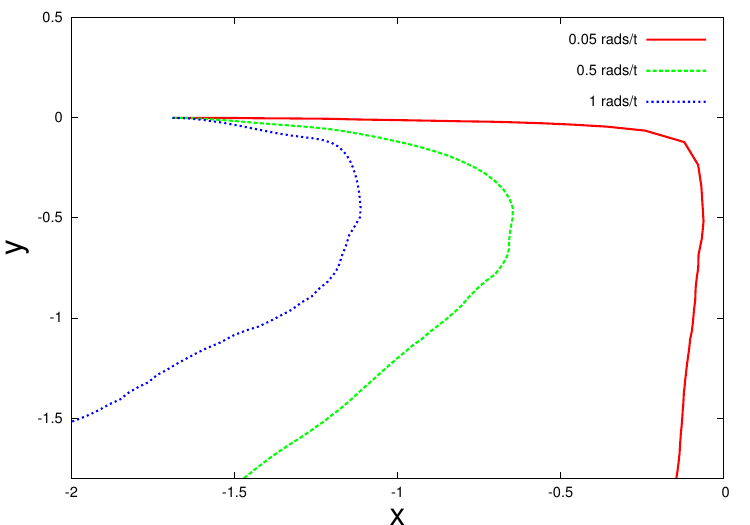}
\caption{Trajectories for spinning scattering \sks of different initial rotational speeds}
\label{trajectories-spinning}
\end{figure}

\Figs \ref{palm-spinning-s}, \ref{palm-spinning-m} and \ref{palm-spinning-f} show the strange effect that the spinning \sks exchange preimages in a spiral pattern, as they scatter. Also, the \sks no longer scatter perpendicularly. This is obvious by the trajectories of the location, shown in \fg \ref{trajectories-spinning}. As the \sks spin faster they deflect more.  These spinning scattering results could help gain a better understanding of the spin-orbit coupling
of nuclei \cite{Kaelbermann:1995ed, Riska:1988zm,Otofuji:1992fw}.

\section{Comparison to monopole scattering}
\label{Monopoles}

In the following, we compare Skyrmion scattering in the attractive channel with monopole scattering. Two-monopole scattering for low velocities can be calculated from geodesic motion in the Atiyah-Hitchin manifold $M_2^0$ \cite{Atiyah}. This four dimensional manifold can be parametrised by a radial coordinate $\rho \in [\frac{\pi}{2},\infty),$  and three angular coordinates $\theta,$ $\phi,$ and $\psi.$ The radial coordinate is basically the separation of the two monopoles, and $\rho=\frac{\pi}{2}$ corresponds to the torus configuration. The angles $\theta$ and $\phi$ parametrise how the monopoles are positioned in ${\mathbb R}^3$ whereas the angle $\psi$ gives the orientation of the monopoles along the axis of separation. The moduli space of monopoles has two important geodesic submanifolds, namely the ``trumpet'' which describes head on collision of monopoles with time-dependent $\psi$, and the ``cone'' which describes monopole scattering in the plane (with $\psi$ constant). We are interested whether there is an analogy of ``rotation without rotating'' in the monopole picture. Skyrmion scattering without rotation in the plane corresponds to monopole scattering along the cone. As we have seen in section \ref{preimages} the effect of ``rotation without rotating'' is related to how much the two Skyrmions overlap. On the monopole moduli space there is a quantity which measures this overlap: the Sen-form \cite{Sen:1994yi} which is exponentially localised at the centre of the monopole moduli space, known as the bolt. 
The hyperk\"ahler $SO(3)$ invariant metric on $M_2^0$ can be written as
\begin{equation}
ds^2 = f^2 d\rho^2 +a^2 {\sigma_1}^2 + b^2 {\sigma_2}^2+c^2 {\sigma_3}^2,
\end{equation}
where $\sigma_k$ are left-invariant one-forms and the coefficient functions satisfy the following differential equations
\begin{eqnarray*}
\frac{2bc}{f}\frac{da}{d\rho} &=&  (b-c)^2-a^2,\\
\frac{2ca}{f}\frac{db}{d\rho} &=&  (c-a)^2-b^2,\\
\frac{2ab}{f}\frac{dc}{d\rho} &=&  (a-b)^2-c^2,
\end{eqnarray*}
where $a(\frac{\pi}{2}) = 0,$ $b(\frac{\pi}{2})=\frac{\pi}{2},$ and 
$c(\frac{\pi}{2})=-\frac{\pi}{2}.$ Here, we follow the conventions in \cite{Gibbons:1986df} and set $f = - b/\rho.$ Then, the Sen form
is the unique normalisable anti-self dual harmonic two-form given by 
\begin{equation}
\omega = F(\rho) \left(d\sigma_1 - \frac{fa}{bc} d\rho \wedge \sigma_1 \right),
\end{equation}
where 
\begin{equation}
F(\rho) = F_0 \exp\left(-\int\limits_{\frac{\pi}{2}}^\rho \frac{fa}{bc} d\rho^\prime\right).
\end{equation}
The Sen form is exact as we can write $\omega = dA$ where $A = F(\rho) \sigma_1.$ Note that $F(\pi) = F_0$ at the bolt. Now, consider a geodesic $\gamma$ in the moduli space $M_2^0.$ Then the path integral
$
\int_\gamma A
$
is equivalent to the loop integral
$
\oint_\gamma A,
$
where we closed the loop via a circle segment at infinity. This does not contribute to the integral due to the asymptotics of $F(\rho),$ namely, $F(\rho)$ is exponentially localised. Using Stokes theorem, 
$$
\oint_\gamma A = \int_D \omega,
$$
where $D$ is the surface bounded by $\gamma.$ This can be interpreted as a holonomy on $M_2^0$ with respect to the Sen form. This holonomy is conjectured to show a very similar behaviour to the ``rotation without rotating'' angle.

\section{Conclusion}

In this paper we discuss Skyrmion-Skyrmion scattering for non-zero impact parameter. Here we focus on the attractive channel where the two Skyrmions are orientated in such a way that the attraction between them is maximal.
For large separation, the scattering can be described in the dipole approximation which ignores the short-range repulsive interaction.  We also discussed the necessary modifications needed to include non-zero pion mass and relativistic corrections. This approximation clearly breaks down at the critical value $b_{{\rm crit}}$ when the two dipoles no longer escape to infinity but collide with each other. For small velocities, Skyrmion scattering in the attractive channel is quantitively similar to monopole scattering which in turn can be described as geodesic motion on the Atiyah-Hitchin manifold. We have calculated Skyrmion trajectories numerically for different velocities and impact parameters, and find good qualitative agreement with the dipole approximation for large impact parameters. 
Note that when Skyrmions are spinning, $\frac{1}{r}$ terms arise
in the Lagrangian \eqref{Lrel} which could have a profound impact on the
dynamics \cite{Schroers:1993yk,Gisiger:1994gj, Gisiger:1994pr,
Gisiger:1998tv}.

When two non-rotating Skyrmions scatter head on, namely with zero impact parameter, in the attractive channel then they scatter by 90 degrees. Using our colouring scheme we observed the following. Initially, the Skyrmions have a relative phase of $\pi.$ During scattering, the Skyrmions move towards each other but do not rotate. Then they form a torus and emerge again from the torus but in a different orientation. While both Skyrmions still have a relative phase of $\pi$ there overall phase has changed by $\frac{\pi}{2}.$ What seemed to have happened is that half of the left Skyrmion has gone up and half of the left Skyrmion has gone down, and similar for the Skyrmion coming from the right. Hence the Skyrmions have rearranged each other, and this leads to a ``rotation without rotating.'' This effect can be explored further by looking at preimages. In a Skyrmion configurations of degree $B=2$ each point in target space generically has at least two preimages. When there are more than two preimages of the same point there has to be negative baryon density, see \cite{Foster:2013bw} for further details. In our simulations, we did not find significants amounts of negative baryon density.
Since for large separations, two Skyrmions are well approximated as hedgehog, we choose the position of the Skyrmions to be $U=-1_2.$ During the scattering process we can generically track the preimages of any point on the sphere and calculate to which final state it belongs. This gives a way of quantifying rotation without rotating, also for non-zero impact parameter. By plotting preimages rather than baryon density we have created a novel way of visualising Skyrmions. 

We also briefly discussed the scattering of spinning Skyrmions. Spinning Skyrmion solutions are not stable for massless pions due to pion radiation. However, we observed pion radiation before the Skyrmions stopped spinning. Spinning Skyrmions no longer scatter at right angles during head-on collision. The configuration of closest approach is also no longer the torus but a configuration which is similar to the stationary solution of isorotation $B=2$ Skyrmions found in \cite{Battye:2014qva}. It would be interesting to compare the dynamics of spinning Skyrmions with the attractive channel approximation in \cite{Schroers:1993yk,Gisiger:1994pr,Gisiger:1998tv}. There has also been recent progress in understanding these classically spinning Skyrmions as approximations to nucleons with quantised spin \cite{Manton:2011mi} and in identifying short-lived resonance states, and also the stable deuteron state, in numerical simulations of scattering events \cite{Foster:2015cpa}.

There are still many open problems in classical Skyrmion-Skyrmion scattering. For example, how do Skyrmions behave for more general initial conditions? 
To what extend can the attractive channel be used to approximate more general scattering events? 
Classically, the Skyrme model is typically too tightly bound. It would be interesting to study Skyrmion scattering in models where the binding energies are lower. This could be achieved by modifying the potential \cite{Gillard:2015eia},  by inclusion of vector mesons \cite{Sutcliffe:2010et} or the BPS part of the Skyrme model \cite{Adam:2010fg}. In fact, an interesting study of scattering trajectories in a model where the Skyrme term has been replaced by coupling pions to omega mesons has been performed in \cite{Amado:1999zr}. Skyrmion scattering in hyperbolic space has recently been studied in \cite{Winyard:2015ula}.
Scattering for higher charges is also an interesting topic. Our preimage technique could provide novel insights into what happens to an individual Skyrmion during scattering. 

While we are currently studying classical scattering, our long-term goal is to understand scattering of nucleons or even the scattering of nuclei. Braaten has outlined how to calculate scattering cross sections in the Skyrme model \cite{Braaten:1987hk}. Gisiger and Paranjape performed analytic calculations of nucleon-nucleon scattering based on the geodesic approximation \cite{Gisiger:1998tv}. We intend to combine these approaches with our scattering results to model experimental data.

\section*{Acknowledgements}

The authors are grateful for fruitful discussions with Nick Manton at various stages of the project. We would also like to thank Mareike Haberichter for useful discussions. SK enjoyed stimulating discussions with Manu Paranjape. This work was financially supported by the U.K. Engineering and Physical Science Research Council (Grant No. EP/I034491/1). DF acknowledges the Leverhulme Trust for financial support as part of the Program Grant: Scientific Properties Of Complex Knots.

\bibliographystyle{../myordered1}
\renewcommand{\baselinestretch}{1}
\begin{small}
\bibliography{../scattering}
\end{small}
\label{lastref}
\end{document}